%% file: proceedings.tex
\documentclass{PoS}

\newcommand{\mydelta}{\Delta}
\newcommand{\lQED}{{\mathrm{QED}}}
\newcommand{\lQCD}{{\mathrm{QCD}}}

\input{newcommand}

\title{Isospin Breaking Effects on the Lattice}

\ShortTitle{Isospin Breaking Effects on the Lattice}

\author{\speaker{Nazario Tantalo}\\
        Universit\`a degli Studi di Roma ``Tor Vergata''\\
        INFN sezione di Roma ``Tor Vergata''\\
        E-mail: \email{nazario.tantalo@roma2.infn.it}}

\abstract{
Isospin symmetry is not exact and the corrections to the isosymmetric limit are, in general, at the percent level. For gold plated quantities, such as pseudoscalar meson masses or the kaon leptonic and semileptonic decay rates, these effects are of the same order of magnitude of the errors quoted in nowadays lattice calculations and cannot be neglected any longer. In this talk I discuss the methods that have been developed in the last few years to calculate isospin breaking corrections by starting from first principles lattice simulations. In particular, I discuss how to perform a combined QCD+QED lattice simulation and a renormalization prescription to be used in order to separate QCD from QED isospin breaking effects. A brief review of recent lattice results of isospin breaking effects on the hadron spectrum is also included.}

\FullConference{31st International Symposium on Lattice Field Theory - LATTICE 2013\\
		July 29 - August 3, 2013\\
		Mainz, Germany}

\begin{document}

\section{Introduction}
The two lightest quarks, the up and the down, have different masses and different electric charges. Nevertheless, their mass difference is much smaller than a typical hadronic scale ($\Lambda_\lQCD$) and electromagnetic interactions are much weaker than strong interactions\footnote{$\hat m_{u}$ and $\hat m_{d}$ are the up and down renormalized quark masses, $\hat \alpha_{em} \simeq 1/137$ the fine structure constant and $e_f$ the fractional electric charge of the $f$ quark, i.e. $e_{u,c,t}=2/3$ and $e_{d,s,b}=-1/3$.},
\begin{eqnarray}
\frac{\hat m_d - \hat m_u}{\Lambda_\lQCD}\ll 1\; ,
\qquad
(e_u - e_d) e_f\ \hat \alpha_{em} \ll 1 \; .
\end{eqnarray}
For this reason isospin, the group of $SU(2)$ flavour rotations in the up-down space, is a mildly broken symmetry and a very useful theoretical tool. For example, thanks to isospin symmetry hadrons can be classified according to the representations of angular momentum algebra, hadronic scattering processes can be studied separately in different ``isospin channels'', the neutral pion two-point correlator has no disconnected diagrams and, on the algorithmic side, unquenched simulations with light Wilson fermions are possible without reweighting because\footnote{$D[U]$ is the massless Wilson lattice Dirac operator depending on the QCD gauge fields $U_\mu(x)$ and $m_{ud}=(m_u+m_d)/2$ is the average up-down bare quark mass.} 
\begin{eqnarray}
\det\left( D[U] + m_{ud} \right)\ \det\left( D[U]^\dagger + m_{ud} \right) > 0 \; .
\end{eqnarray}

Isospin breaking is a small effect but generates a rich phenomenology, for example chemistry. The hydrogen atom is stable because $M_n-M_p>M_e$ and the electron capture reaction $p+e\mapsto n+\nu_e$ is forbidden. As discussed in the following, the separation of QCD from QED isospin breaking corrections is unphysical and depends upon the renormalization conditions. By choosing a ``natural'' prescription one has that the neutron is heavier than the proton thanks to a delicate balance between two opposite contributions of the same order of magnitude,  $(M_n-M_p)^\lQED<0<(M_n-M_p)^\lQCD$, see Figure~\ref{fig:baryons}. Other interesting examples of phenomena that originate from the breaking of isospin symmetry are the mixings and the decay patterns of neutral mesons or the more recent puzzle of the flavour structure of the ``new'' $X,Y,Z$ hadrons~\cite{Esposito:2013fma}.

In flavour physics there are observables that have been computed on the lattice in the isosymmetric limit with very high accuracy. According to the FLAG2 average~\cite{Aoki:2013ldr}, we know the ratio\footnote{$F_K$ and $F_\pi$ are the kaon and pion decay constants in the isosymmetric limit while $F_+^{K\pi}(q^2)$ is the form factor entering the semileptonic decay rate of a kaon into a pion in the isosymmetric limit ($F_+^{K\pi}=F_+^{K^0\pi^-}=\sqrt{2}F_+^{K^+\pi^0}$).} $F_K/F_\pi$ and the zero recoil form factor $F_+^{K\pi}(0)$ with an accuracy of $\sim 0.4\%$.
QCD isospin breaking effects on these quantities have been estimated in chiral perturbation theory~\cite{Kastner:2008ch,Cirigliano:2011tm} and are expected to be $\sim -0.2\%$ for the ratio of decay constants and as large as $3\%$ for the form factor. We are rapidly approaching a situation in which it will be useless to put efforts in further reducing the uncertainty on isosymmetric hadronic observables if isospin breaking effects (IBE) are not taken into account from first principles.

\section{QCD+QED on the lattice}
The IBE associated with electromagnetic interactions are as important as the effects associated with the up-down mass splitting. This means that in order to have an in impact on phenomenology lattice calculations of IBE require simulations of what we call the full theory\footnote{We call isosymmetric theory QCD with the masses of the up and of the down set equal to the common value $m_{ud}$.}, i.e. QCD+QED. 
Full theory observables are defined in terms of the following path-integral average\footnote{The bare parameters of the full theory (ignoring heavy flavour masses) are collected in the vector $\vec g$; $\beta=6/g_s^2$; $A_\mu(x)$ is the photon field, the dynamical variable in the non-compact formulation of QED (see below); $D[U,A;\vec g]$ is the preferred discretization of the Dirac operator.}
\begin{eqnarray}
\vec g = \Big(e^2,g_s^2,m_{u},m_{d},m_s\Big) \; ,
\qquad
\langle O \rangle^{\vec g} =
\frac{
\int{
dAe^{-S[A]}\; dU\; e^{-\beta S[U]}\; \prod_f
\det\left(D_f[U,A;\vec g]\right)\; O[U,A;\vec g]
}}
{
\int{
dAe^{-S[A]}\; dU\; e^{-\beta S[U]}\; \prod_f
\det\left(D_f[U,A;\vec g]\right)
}} \; .
\nonumber \\
\end{eqnarray}
The direct generation of QCD+QED gauge configurations is possible, in principle, with lattice fermion actions such that the determinant of the single flavour is real and positive-definite. In practice this procedure would be too much expensive or at least unpractical. It is much more efficient to re-use the gauge configurations generated in the isosymmetric theory\footnote{The vector $\vec g^0$ collects the bare parameters of the isosymmetric QCD.},
\begin{eqnarray}
\vec g^0=\Big(0,(g_s^0)^2,m_{ud}^0,m_{ud}^0,m_s^0\Big) \; ,
\qquad
\langle O \rangle^{\vec g^0} =
\frac{
\int{
dU\; e^{-\beta^0 S[U]}\; \prod_f
\det\left(D_f[U;\vec g^0]\right)\;  O[U;\vec g^0]
}}
{
\int{
dU\; e^{-\beta^0 S[U]}\; \prod_f
\det\left(D_f[U;\vec g^0]\right)
}} \; .
\nonumber \\
\end{eqnarray}
This can be done by introducing the ``QED path-integral average'' and a reweighting factor
\begin{eqnarray}
\langle O \rangle^A = 
\frac{\int{dA\; e^{-S[A]}\; O[A]}}
{\int{dA\; e^{-S[A]}}} \; ,
\qquad
R[U,A;\vec g,\vec g^0] &=& e^{-(\beta-\beta^0) S[U]}\ \prod_f
\frac{\det\left(D_f[U,A;\vec g]\right)}{\det\left(D_f[U;\vec g^0]\right)} \; ,
\label{eq:reweightfactor}
\end{eqnarray}
and by writing $\langle O \rangle^{\vec g}$ as follows
\begin{eqnarray}
\langle O \rangle^{\vec g} =
\frac{\big\langle\; R[U,A;\vec g,\vec g^0]\; 
O[U,A;\vec g] \; \big\rangle^{A,\vec g^0}}
{\big\langle\; R[U,A;\vec g,\vec g^0]\; \big\rangle^{A,\vec g^0}} \; .
\label{eq:tobeexpanded}
\end{eqnarray}
The formulae above and the numerical calculations are much more simple in the so-called ``electroquenched'' approximation, i.e. by considering sea quarks as electrically neutral particles. This ``rough'' approximation leads to a non-unitary theory and is obtained by setting
\begin{eqnarray}
R[U,A;\vec g,\vec g^0] & \mapsto& 1 \; .
\end{eqnarray}
Electroquenched QED ensembles can be obtained easily and efficiently with heat-bath algorithms.

The first pioneering lattice calculation of IBE has been performed in ref.~\cite{Duncan:1996xy} by relying on the electroquenched approximation. In that reference and also in the more recent works on the subject QED has been simulated in the non-compact formulation: the gauge potential $A_\mu(x)$ is a dynamical variable and the QCD+QED links are obtained by exponentiation,
\begin{eqnarray}
U_\mu(x) &\mapsto& e^{ie_f e A_\mu(x)}\ U_\mu(x) \; .
\end{eqnarray}
Imposing periodic boundary conditions for the gauge potential and a gauge fixing (here Feynman),
\begin{eqnarray}
\nabla^-_\mu A_\mu(x)=0 \; ,
\qquad
S[A]
=
\frac{1}{2}\sum_{x}{
A_\mu(x)\left[-\nabla^-_\nu\nabla^+_\nu\right]A_\mu(x)} \; ,
\end{eqnarray}
the QED gauge action has a zero mode and the photon propagator is \emph{infrared divergent}. Furthermore, the Guass law is inconsistent (see for example ref.~\cite{Hayakawa:2008an}). Both problems are solved by subtracting the zero momentum mode, a residual gauge ambiguity associated with any derivative gauge fixing,
\begin{eqnarray}
0=\nabla^-_\mu A_\mu(x)= \nabla^-_\mu \left[ A_\mu(x) + c\right] \;.
\end{eqnarray}
It can be shown that this infrared regularization changes physical quantities by finite volume effects, there are no new ultraviolet divergences to cope with. Note that QED is a long range unconfined interaction and (large) finite volume effects are unavoidable. The infrared regularized QED action can be written directly in coordinate space, without the need of (fast) Fourier transforms, by introducing a suitable projector~\cite{deDivitiis:2013xla}
\begin{eqnarray}
\mathtt{P^\perp} \phi(x)=\phi(x) - \frac{1}{V}\sum_y \phi(y) \; ,
\qquad
S[A]
\mapsto
\frac{1}{2}\sum_{x}{
A_\mu(x)\left[-\nabla^-_\nu\nabla^+_\nu\right]\mathtt{P^\perp}\ A_\mu(x)} \; .
\label{eq:regphotonaction}
\end{eqnarray}
\begin{figure}[!t]
\begin{center}
\includegraphics[width=0.49\textwidth]{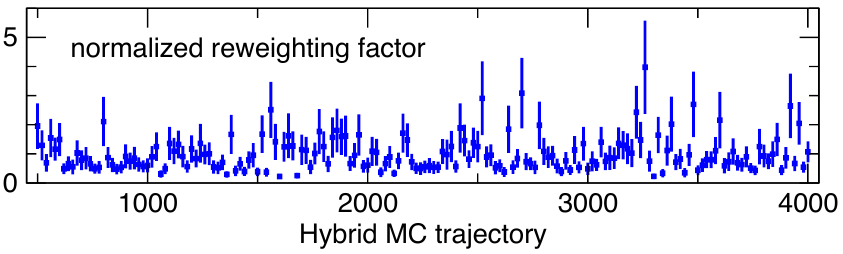}\hfill
\includegraphics[width=0.49\textwidth]{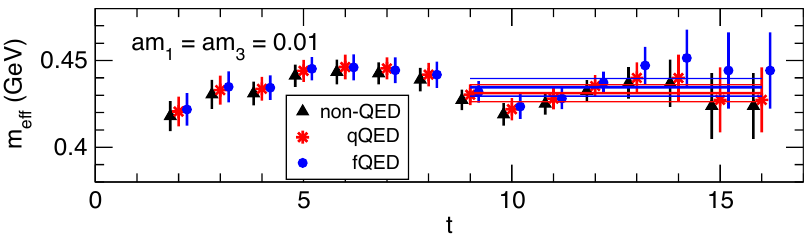}\hfill
\caption{\label{fig:rewfluctuations} \footnotesize
Left: fluctuations of the reweighting factor. Right: effective mass of a pseudoscalar meson extracted on the same QCD gauge background in the isosymmetric theory (black), in the electroquenched theory (red) and in the full theory (blue). Both the figures are taken from Ishikawa et al.~\cite{Ishikawa:2012ix}, see this reference for further details.
}  
\end{center}
\end{figure}
Recently Ishikawa et al.~\cite{Ishikawa:2012ix} and the PACS-CS collaboration~\cite{Aoki:2012st} have demonstrated the feasibility of simulations of the full theory beyond the electroquenched approximation. In both these works the physical volumes are of the order of $3$~fm and the reweighting factor, see eq.~(\ref{eq:reweightfactor}), has been split into several factors with controllable statistical fluctuations. Ishikawa et al. factored $R$ by using the $n^{th}$-root trick while the PACS-CS collaboration used a mass-charge preconditioning. The plots in Figure~\ref{fig:rewfluctuations} are taken from ref.~\cite{Ishikawa:2012ix} but similar plots can be found in ref.~\cite{Aoki:2012st} (see also ref.~\cite{Finkenrath:2013soa}).
In the left panel it is shown the HMC history of the reweighting factor normalized by its average. 

When QED interactions are introduced through reweighting and simulations are performed at the physical value of the electric charge the resulting IBE are typically smaller than statistical errors, see the right panel in Figure~\ref{fig:rewfluctuations}. In ref.~\cite{Ishikawa:2012ix} it is observed that IBE can however be calculated by relying on the strong statistical correlations between the different data sets (black, red and blue) that share the same QCD gauge background. In fact physics is associated with the full theory and, although interesting and possibly convenient from the numerical point of view, there is no need to consider the difference between isosymmetric and full theory results. This is an important and subtle point that we are now going to discuss in some detail.

\section{Calibration of the lattice: QCD vs. QCD+QED}
QCD+QED and QCD are two different theories. Electromagnetic currents generate divergent contributions,
\begin{eqnarray}
(e_f  e)^2 \golself \quad &\longrightarrow&\quad [m_f-m_f^0] \goi \;,
\nonumber \\
\nonumber \\
J^\mu(x)J_\mu(0)
\quad&\longrightarrow&\quad
c_1(x) \mathtt{1} + 
\sum_f \left[c_m^f(x)  m_f + c_{cr}^f(x) \right]
\bar \psi_f \psi_f +
c_{g_s}(x) 
G_{\mu\nu}G^{\mu\nu}
+\cdots \; ,
\label{eq:opejj}
\end{eqnarray}
that redefine the vacuum energy, $c_1$, the quark masses, $c_m^f$, the quark critical masses (if chirality is broken), $c_{cr}^f$, and the strong coupling constant (the lattice spacing), $c_g$. The parameters of the physical theory, QCD+QED, can be fixed by using a suitable number of experimental inputs. This is the approach followed by the PACS-CS collaboration in ref.~\cite{Aoki:2012st} where the experimental determinations of $\{ M_{\pi^+},M_{K^+},M_{K^0},M_{\Omega^-}\}$ have been used to tune $\{ \hat m_u,\hat m_d,\hat m_s,a\}$ and, of course, the masses of the up and of the down turned out to be different. That's it.

On the other hand it is theoretically interesting and possibly numerical convenient to define differences as $M_H^{\lQCD+\lQED}-M_H^{\lQCD}$ where $M_H$ is the mass of a generic hadron. To this end the ``unphysical'' parameters of the isosymmetric theory have to be set by giving a renormalization prescription. A possibility is to use an hadronic scheme in both theories. One could for example perform a ``standard'' QCD simulation and use $\{ M_{\pi^+},M_{K^+},M_{\Omega^-}\}$ to fix $\{ \hat m_{ud}^0,\hat m_s^0,a^0\}$. If the parameters of the full theory are then fixed as done by the PACS-CS collaboration, there would be no IBE on $\{ M_{\pi^+},M_{K^+},M_{\Omega^-}\}$ in this scheme while IBE could be properly defined and calculated for any other observable.

In ref.~\cite{deDivitiis:2013xla}, see also ref.~\cite{Gasser:2003hk}, it has been suggested to define IBE by using an intermediate renormalization scheme and a matching procedure. To implement this prescription one has to: tune the full theory bare parameters $g_i$ by using experimental inputs; choose a renormalization scheme ($\overline{MS}$ or a non-perturbative scheme as SF or RI-MOM) and a matching scale $\mu^\star$; fix the renormalized parameters of the isosymmetric theory ($\hat \alpha_{em}=\hat m_d-\hat m_u =0$) by the matching condition $\hat g_i^0(\mu^\star)=\hat g_i(\mu^\star)$. Note that the renormalized parameters of the two theories, although equal in this scheme at the scale $\mu^\star$, are different at any other scale. Naturally, also the bare parameters are different\footnote{$Z_i(\mu)$ are the renormalization constants of the full theory, $\hat g_i= Z_i(\mu) g_i$, while $Z_i^0(\mu)$ are the renormalization constant of isosymmetric QCD, $\hat g_i^0= Z_i^0(\mu) g_i^0$.}
\begin{eqnarray}
g_i^0 = \frac{Z_i(\mu^\star)}{Z_i^0(\mu^\star)} g_i \; .
\end{eqnarray}
Once the parameters have been fixed, IBE for any observable can be properly defined as 
\begin{eqnarray}
\mydelta O=O(\vec g)-O(\vec g^0) \; .
\label{eq:ibedef}
\end{eqnarray}
A similar procedure can be used for instance to properly define unquenching effects and to compare $n_f=2+x$ with $n_f=2+y$ lattice results.

In the case of light pseudoscalar meson observables, the matching of QCD+QED with QCD can be performed by fitting lattice results to analytical formulae derived in chiral perturbation theory coupled to electromagnetic interactions~\cite{Bijnens:2006mk,Hayakawa:2008an}. All the terms allowed by symmetries are present in the chiral formulae that can be expressed either in terms of the renormalized parameters of the full theory or, by a redefinition of the low energy constants, in terms of the renormalized couplings of isosymmetric QCD. This is the strategy followed in refs.~\cite{Basak:2013iw,Ishikawa:2012ix,Blum:2010ym} and in previous works on the subject. Although the matching is somehow ``automatic'' in this approach, the details of the renormalization prescriptions have to be specified when quoting results to allow their comparison with other determinations and with experimental data.

In the following we shall talk about ``leading isospin breaking effects'' (LIBE). These are defined by expanding eq.~(\ref{eq:ibedef}) in powers of\footnote{Note the absence in eq.~(\ref{eq:expbare}) of terms linear in $e$ and $g_s$ (physical observables are QED and QCD gauge invariant) and the presence of a term proportional to the shift of the critical masses $m_f^{cr}-m_0^{cr}$ that is needed in theories in which chirality is broken.} $g_i-g_i^0$,
\begin{eqnarray}
\mydelta O
&=&
\left\{
e^2 \frac{\partial}{\partial e^2} +
\left[g_s^2- (g_s^0)^2\right] 
\frac{\partial}{\partial g_s^2}+
[m_f-m_f^0] \frac{\partial}{\partial m_f}+
[m_f^{cr}-m_0^{cr}] \frac{\partial}{\partial m_f^{cr}}
\right\} O \; .
\label{eq:expbare}
\end{eqnarray}
Note that the counter-terms in the perturbative expansion with respect to $\hat \alpha_{em}$, i.e. in the operator product expansion of eq.~(\ref{eq:opejj}), do arise because the bare parameters (the renormalization constants) of the two theories are different. Indeed, once expressed in terms of renormalized quantities, eq.~(\ref{eq:expbare}) becomes
\begin{eqnarray}
\mydelta O
&=&
\left\{
\hat e^2 \frac{\partial}{\partial \hat e^2} +
\left[\hat g_s^2- \left(\frac{Z_{g_s}}{Z^0_{g_s}}
\hat g_s^0\right)^2\right]
\frac{\partial}{\partial \hat g_s^2}+
\left[\hat m_f-\frac{Z_{m_f}}{Z^0_{m_f}}\hat m_f^0\right]
\frac{\partial}{\partial \hat m_f}+
\Delta m^{cr}_f \frac{\partial}{\partial m^{cr}_f}
\right\}O \; .
\end{eqnarray}
The divergent quantities $Z_{m_f}/Z^0_{m_f}$,
$\Delta m^{cr}_f=m_f^{cr}-m_0^{cr}$ and $Z_{g_s}/Z^0_{g_s}$ appearing in the previous equation correspond to the counter-terms $c_m^f$, $c_{cr}^f$ and $c_{g_s}$ of eq.~(\ref{eq:opejj}). The electric charge does not need to be renormalized at this order,
\begin{eqnarray}
\hat e^2 = e^2 = 4\pi \hat \alpha_{em} = \frac{4\pi}{137} \;,
\end{eqnarray}
The problem of the renormalization of the electric charge would have to be faced in the calculation of next-to-leading IBE. From the phenomenological point of view, given the size of the other hadronic uncertainties, sub-leading IBE can be safely neglected by now. Note that whenever lattice data are analyzed by neglecting terms of $O[\hat \alpha_{em}(\hat m_d-\hat m_u)]$ one is actually computing LIBE.  

\section{LIBE as a perturbation}
\begin{figure}[!t]
\begin{center}
\includegraphics[width=0.8\textwidth]{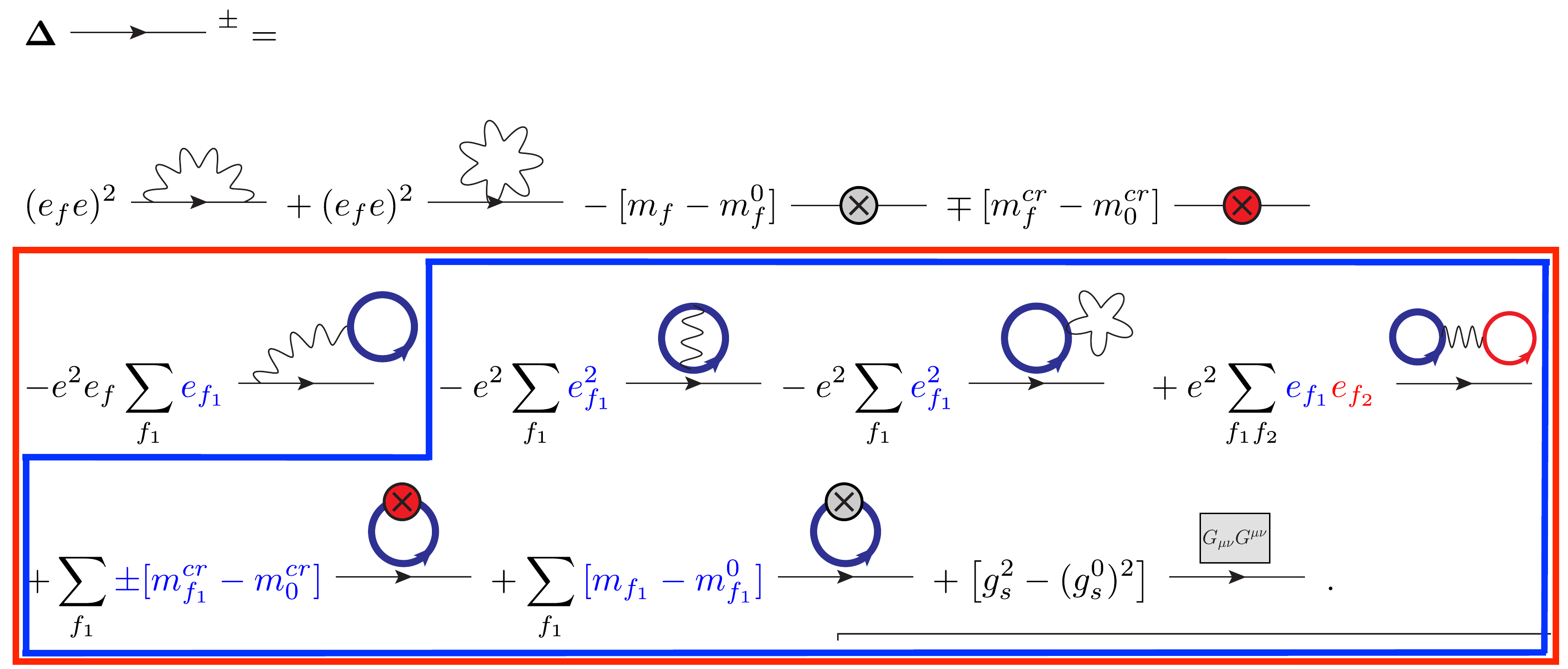}
\caption{\label{fig:graphnotation} \footnotesize
LIBE corrections to the quark propagator (at fixed gauge QCD background) in the graphical notation of ref.~\cite{deDivitiis:2013xla}. The contributions contained in the red box are absent in the electroquenched approximation. The contributions contained in the blue box do not ``read'' the charge of the valence quarks and are therefore isosymmetric.
}  
\end{center}
\end{figure}
In refs.~\cite{deDivitiis:2011eh,deDivitiis:2013xla} it has been shown that LIBE can be calculated efficiently and accurately by expanding the lattice QCD+QED path-integral of eq.~(\ref{eq:tobeexpanded}) in powers of $g_i-g_i^0$
\begin{eqnarray}
O({\vec g}) &=&
\frac{\big\langle\; \left( 1 + \dot{R}+\cdots \right)\; 
\left(O + \dot{O}+\cdots \right)\; \big\rangle^{A,\vec g^0}}
{\big\langle\;  1 + \dot{R} +\cdots\; \big\rangle^{A,\vec g^0}}
\ =\ 
O({\vec g}^0) + \mydelta O \; .
\end{eqnarray}
In these references it has been developed a ``graphical notation'' as a tool to make calculations. The building blocks of the graphical notation are the corrections to the quark propagator (at fixed QCD gauge background) shown in Figure~\ref{fig:graphnotation}. A dictionary to translate in local operator language the different graphical contributions can be found in ref.~\cite{deDivitiis:2013xla}. The contributions of Figure~\ref{fig:graphnotation} contained in the red box are absent in the electroquenched approximation. The ``isosymmetric vacuum polarization'' terms, those contained in the blue box, do not ``read'' the charge of the valence quarks and are expected to be sizeable (see ref.~\cite{Ishikawa:2012ix} for a first numerical evidence). The polarization effects proportional to the charges of the valence quarks are a flavour $SU(3)$ breaking effect. In the case of pseudoscalar meson masses these can be estimated by the knowledge of the low energy constants entering the leading order chiral perturbation theory lagrangian in presence of electromagnetic interactions~\cite{Bijnens:2006mk}.

The starting point of the calculation of LIBE on the mass of a given hadron $H$ is the full theory two-point correlator
\begin{eqnarray}
C_{HH}(t;\vec g)=\langle\ O_H(t)\ O_H^\dagger(0)\ \rangle^{\vec g}
\quad \longrightarrow \quad
e^{M_H} = \frac{C_{HH}(t-1;\vec g)}{C_{HH}(t;\vec g)}+ \ \mbox{non leading exps.} \; ,
\end{eqnarray}
where $O_H$ is an interpolating operator with the quantum numbers of $H$. If $H$ is a charged particle, the correlator $C_{HH}$ is \emph{not} QED gauge invariant. For this reason it is not possible, in general, to extract physical information directly from the residues of the different poles. 
\begin{figure}[!t]
\begin{center}
\includegraphics[width=0.2\textwidth]{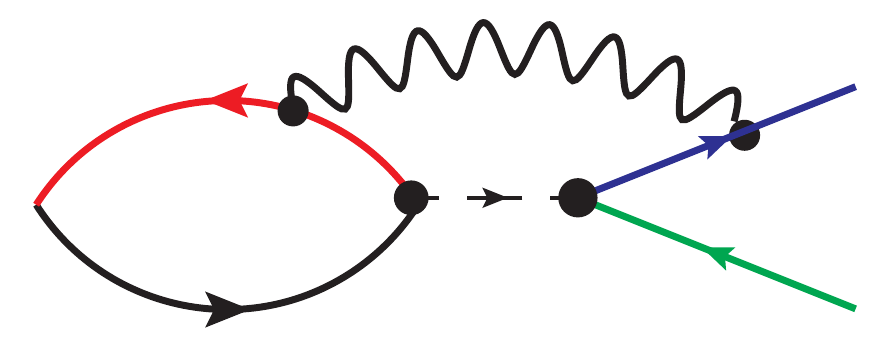}
\caption{\label{fig:exchange} \footnotesize
Example of a non factorable diagram contributing to the physical leptonic decay rate at $O(\hat \alpha_{em})$. In general, the sum of factorable contributions is not QED gauge invariant, infrared divergent and, consequently, unphysical.
}  
\end{center}
\end{figure}
This can be understood by noting that to physical decay rates contribute diagrams as the one shown in Figure~\ref{fig:exchange}. On the other hand, the mass of the hadron is gauge invariant and, provided that the parameters of the action have been properly renormalized, both ultraviolet and infrared finite. It follows that (for large times) the ratio $C_{HH}(t-1;\vec g)/C_{HH}(t;\vec g)$ is both gauge and renormalization group (RGI) invariant. By expanding the numerator and the denominator of this ratio one gets a formula for LIBE on hadron masses,
\begin{eqnarray}
&&\frac{C_{HH}(t;\vec g)}{C_{HH}(t;\vec g^0)} =
   1 + \frac{\mydelta C_{HH}(t;\vec g^0)}{C_{HH}(t;\vec g^0)}
  +\cdots \ =\ c -t(M_H -M_H^0)+\dots \; ,
\nonumber \\
\nonumber \\
&&-\partial_t\frac{\mydelta C_{HH}(t;\vec g^0)}{C_{HH}(t;\vec g^0)}
+\cdots \ =\ M_H -M_H^0 \; .
\label{eq:slopes}
\end{eqnarray}

The pion mass splitting is a particularly ``clean'' observable. In ref.~\cite{deDivitiis:2013xla} it has been derived the elegant formula
\begin{eqnarray}
M_{\pi^+}-M_{\pi^0}
&=&
\frac{(e_u-e_d)^2}{2} e^2\partial_t{\color{red}}\frac{\gdllexch-\discgdllexch}{\gdll}\; .
\label{eq:pionmasses}
\end{eqnarray}
Note: there are no corrections proportional to $\hat m_d-\hat m_u$, i.e. the pion mass difference at this order is a pure electromagnetic effect; vacuum polarization effects are the same for $M_{\pi^+}$ and $M_{\pi^0}$ and cancel exactly in the difference; $M_{\pi^+}-M_{\pi^0}$ is a genuine isospin breaking effect and, for this reason, the electromagnetic shift of the lattice spacing enters at higher orders; since also the electric charge does not renormalize at this order, eq.~(\ref{eq:pionmasses}) is ultraviolet finite. 

The fermion disconnected diagram appearing in eq.~(\ref{eq:pionmasses}) has been neglected, to my knowledge, in all the numerical calculations performed so far. Actually it can be shown, see ref.~\cite{deDivitiis:2013xla}, that this is an $O(\hat m_{ud} \hat \alpha_{em})$ effect and, for physical values of the average up-down mass, it can be considered of the same order of magnitude of next-to-leading IBE. The remaining contribution, the ``exchange'' diagram, can be calculated as an isosymmetric QCD observables by the following procedure. Introducing a real $\mathbb{Z}_2$ noise,
\begin{eqnarray}
\left\langle B_\mu(x) B_\nu(y) \right\rangle^B =\delta_{\mu\nu}\ \delta(x-y) \;, 
\end{eqnarray}
the infrared regularized photon propagator can be calculated by solving
\begin{eqnarray}
[-\nabla^-_\rho\nabla^+_\rho] C_\mu[B;x] = \mathtt{P^\perp}\; B_\mu(x) \; ,
\end{eqnarray}
where $\mathtt{P^\perp}$ has been defined in eq.~(\ref{eq:regphotonaction}). The calculation of the exchange diagram can thus be reduced to two sequential quark propagator inversions,
\begin{eqnarray}
\left\{ D_f[U]\; \Psi_B^f \right\}(x) &=& \sum_\mu B_\mu(x) \Gamma_V^\mu S_f[U;x]  \; ,
\nonumber \\
\nonumber \\
\left\{ D_f[U]\; \Psi_C^f\right\} (x) &=& \sum_\mu C_\mu[B;x] \Gamma_V^\mu S_f[U;x]  \; ,
\end{eqnarray}
where $\Gamma_V^\mu$ is the lattice quark-photon-quark vertex, a functional of the QCD gauge background. We get
\begin{eqnarray}
\gdllexch = \left\langle\ \mbox{Tr}\left\{\ [\Psi_C^{ud}]^\dagger(t)\ 
\Psi_B^{ud}(t)\ \right\} \ \right\rangle^{B}
\; .
\end{eqnarray}
\begin{figure}[!t]
\begin{center}
\includegraphics[width=0.49\textwidth]{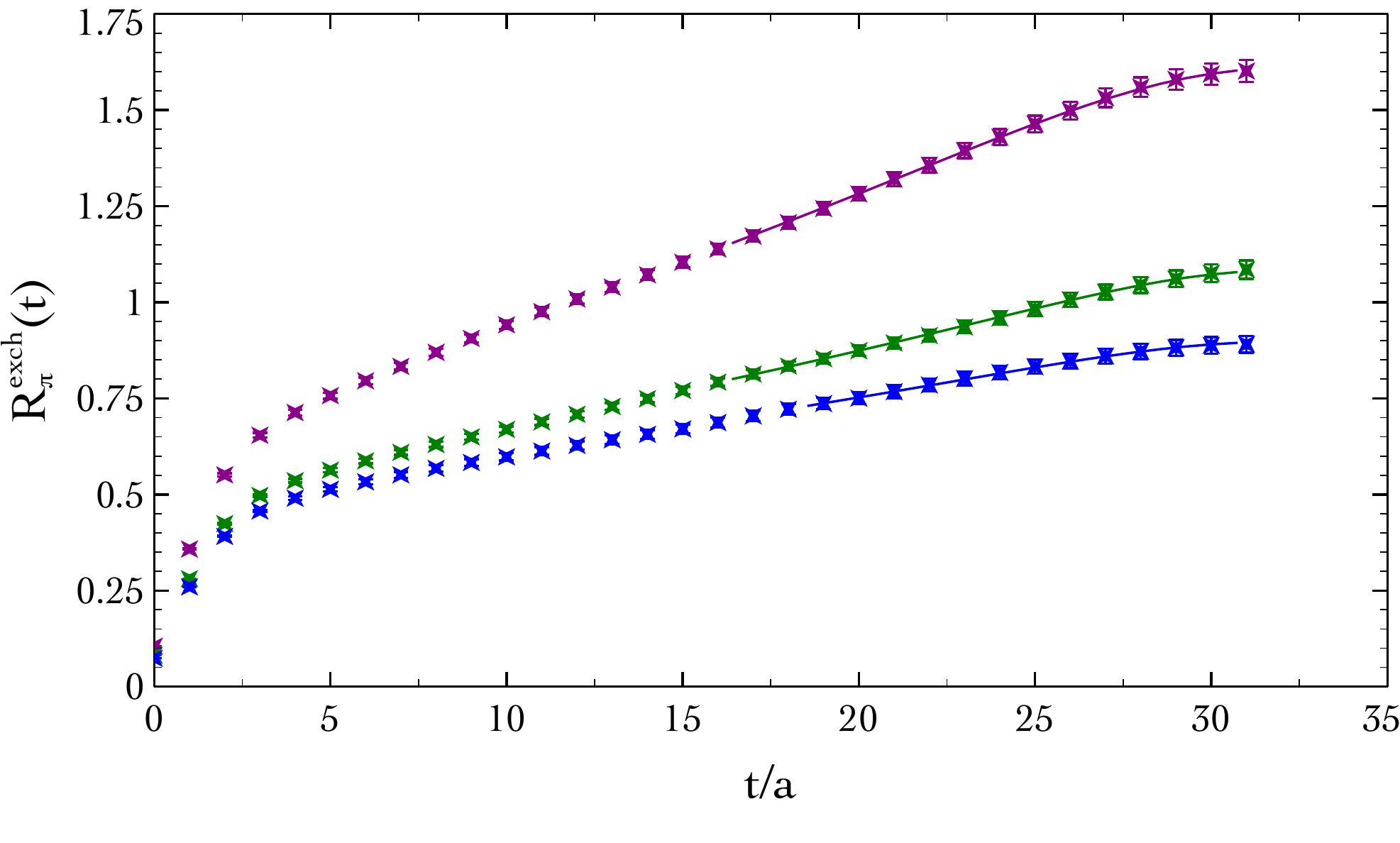}
\includegraphics[width=0.49\textwidth]{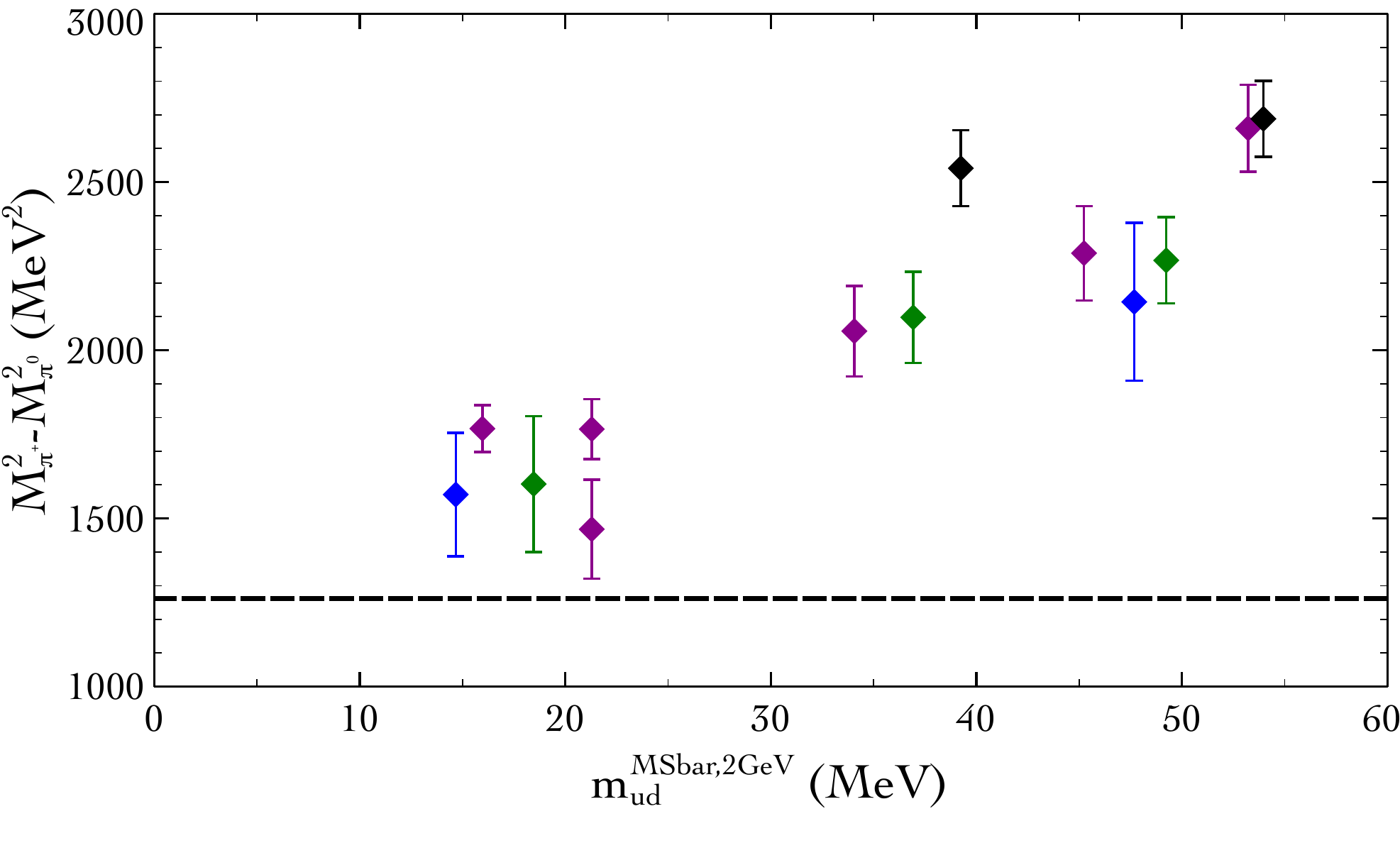}
\caption{\label{fig:ourslopes} \footnotesize
Numerical results obtained in ref.~\cite{deDivitiis:2013xla} for the direct computation of LIBE on pion masses. Different colours correspond to different (black coarser, blue finest) lattice spacings. Finite volume corrections, which are not negligible (see discussion below), have not been taken into account yet in the plot.
}  
\end{center}
\end{figure}
Figure~\ref{fig:ourslopes} shows the results obtained in ref.~\cite{deDivitiis:2013xla} for the pion mass splitting by neglecting the fermion disconnected diagram in eq.~(\ref{eq:pionmasses}). The different data sets correspond to different lattice spacings. The results for $M_{\pi^+}-M_{\pi^0}$ shown in the right panel are obtained by taking the derivative with respect to the time of the correlators in the left panel of the Figure. 
By comparing the left panel of Figure~\ref{fig:ourslopes} with the right panel of Figure~\ref{fig:rewfluctuations} one can appreciate the quality of the numerical signals usually obtained in direct calculations of LIBE. The point is that IBE are tiny because very small coefficients multiply sizeable hadronic matrix elements. 
On the other hand, the direct approach to LIBE requires in general the calculation of several contributions, see next section.

\section{Separation of QCD from QED IBE}
In the graphical notation of ref.~\cite{deDivitiis:2013xla} the kaon mass splitting is given by
\begin{eqnarray}
M_{K^+}-M_{K^0}&=&
-2{\color{red}}\Delta  m_{ud} \partial_t \frac{\gdsi}{\gdsl}
-({\color{blue}} \Delta m^{cr}_u-\Delta m^{cr}_d)\partial_t \frac{\gdsip}{\gdsl}
\nonumber \\
&+&(e_u^2-e_d^2)  e^2 \partial_t \frac{\gdslexch-\gdslselfl-\gdslphtadl}{\gdsl}
+(e_u-e_d)  e^2 \sum_f{ e_f \partial_t \frac{\gdslltadf}{\gdsl}} \;.
\label{eq:kaonmasses}
\end{eqnarray}
The contributions in the first line of the previous equation are the mass and critical mass counter-terms. Whenever electromagnetic ``self-energy'' contributions are present, as in the second line of eq.~(\ref{eq:kaonmasses}), the mass counter-terms are also present because these are needed to absorb the electromagnetic ultraviolet divergences. 

Given the presence of the term proportional to $\Delta m_{ud}= (m_d-m_u)/2$, the kaon mass splitting can be used to determine the up-down mass difference and to define a prescription to separate QCD from QED IBE. First note that since $e_u\neq e_d$ there is a mixing in the renormalization of the full theory between the parameters $\Delta \hat m_{ud}$ and $\hat m_{ud}$,
\begin{eqnarray}
\Delta m_{ud}
\quad =\quad  \frac{\hat m_d}{2Z_{m_d}}-\frac{\hat m_u}{2Z_{m_u}}
\quad=\quad Z_{\bar \psi \psi} \Delta \hat m_{ud} + 
\frac{\hat m_{ud}}{\mathcal{Z}_{ud}} \; .
\end{eqnarray}
The renormalization constant $Z_{\bar \psi \psi}=1/2Z_{m_d}+1/2Z_{m_u}$ has to be replaced with the renormalization constant $Z_{\bar \psi \psi}^0=1/Z_m$ of isosymmetric QCD while, to a first approximation, $\mathcal{Z}_{ud}$ can be safely calculated in perturbation theory,
\begin{eqnarray}
\frac{1}{\mathcal{Z}_{ud}}\quad=\quad\frac{1}{2Z_{m_d}}-\frac{1}{2Z_{m_u}}
\qquad \longrightarrow \qquad
\frac{(e_d^2-e_u^2)e^2}{32\pi^2}\left[
\gamma_{\bar \psi \psi}\log(a\mu^\star) + \mbox{ finite }
\right] Z_{\bar \psi \psi}^0 \; .
\end{eqnarray}

A convenient prescription to separate QCD from QED IBE is given by
\begin{eqnarray}
&&\left[M_{K^+}-M_{K^0}\right]^\lQED(\mu^\star)=
\nonumber \\
&&  -{\color{red}}\frac{2\hat m_{ud}}{\mathcal{Z}_{ud}}
  \partial_t \frac{\gdsi}{\gdsl}
  -(\Delta m^{cr}_u-\Delta m^{cr}_d) \partial_t \frac{\gdsip}{\gdsl}
  +
  (e_u^2-e_d^2)  e^2\partial_t \frac{\gdslexch-\gdslselfl-\gdslphtadl}{\gdsl} \; ,
\nonumber 
\nonumber \\
&&\left[M_{K^+}-M_{K^0}\right]^\lQCD(\mu^\star)=
  -{\color{red}}2\Delta  \hat m_{ud}
  \left( Z_{\bar \psi \psi}^0\ \partial_t \frac{\gdsi}{\gdsl} \right) \; .
\label{eq:QCDkaons}
\end{eqnarray}
All the terms appearing in $\left[M_{K^+}-M_{K^0}\right]^\lQED$ vanish if the electric charges of the up and of the down are taken equal. Furthermore, the definition of $\left[M_{K^+}-M_{K^0}\right]^\lQCD$ is RGI invariant in the isosymmetric theory, $Z_{\bar \psi \psi}^0(\mu^\star)\ \Delta \hat m_{ud}(\mu^\star) = \Delta m_{ud}^0$.
Once a simulation of the full theory has been performed and a value of $\left[M_{K^+}-M_{K^0}\right]^\lQCD$ has been obtained, this can be used as the ``experimental'' input needed in non isosymmetric QCD simulations to tune the up-down mass difference.

\begin{figure}[!t]
\begin{center}
\includegraphics[width=0.49\textwidth]{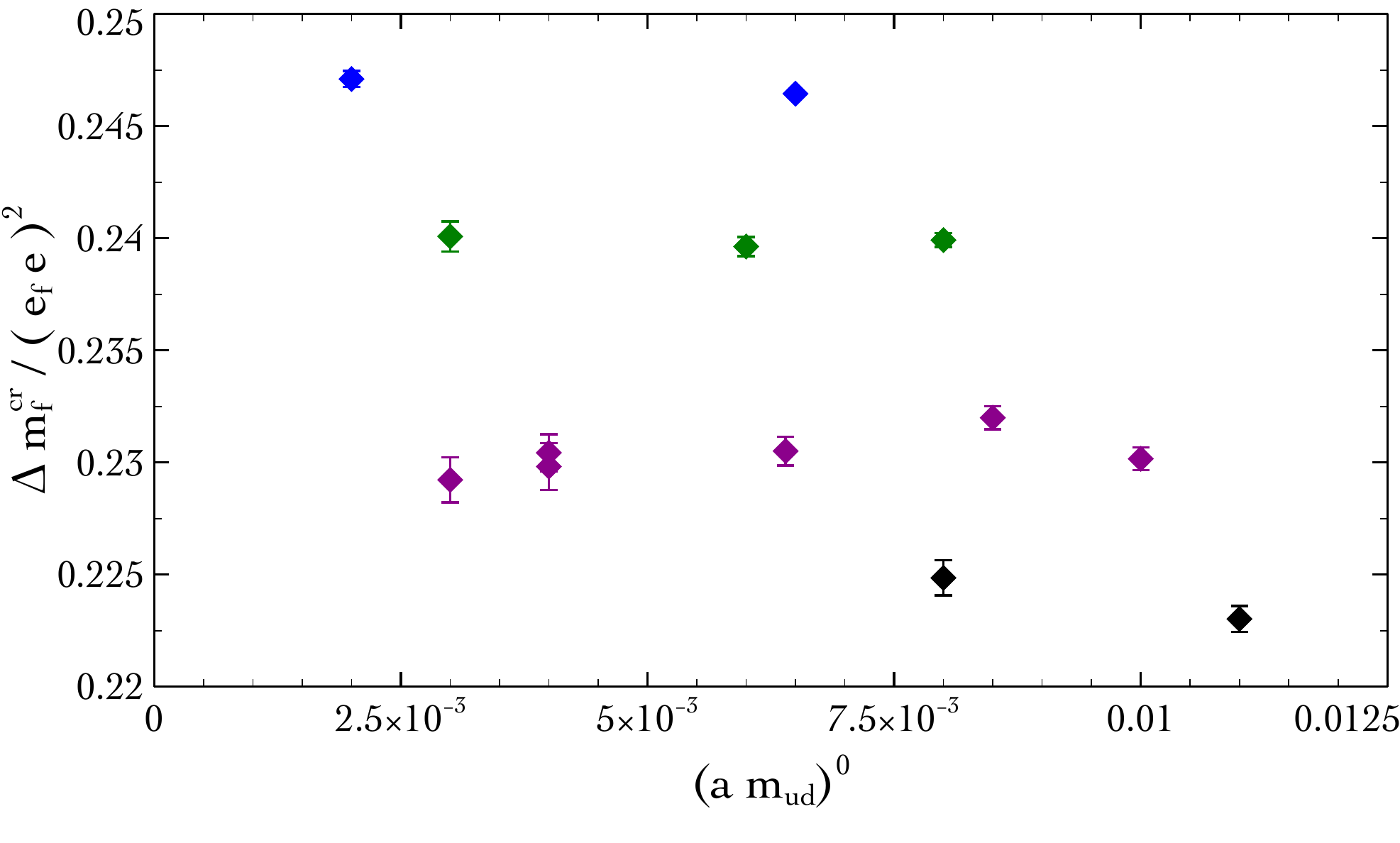} \hfill
\includegraphics[width=0.4\textwidth]{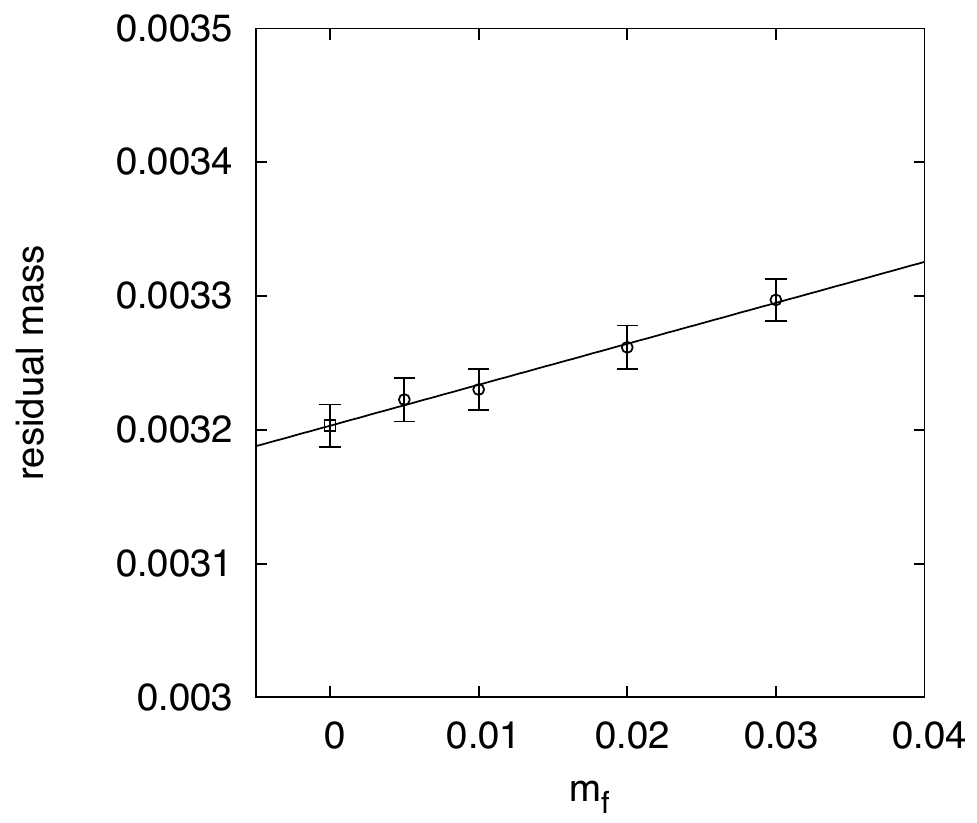}
\caption{\label{fig:mcritical} \footnotesize
Tuning of the critical mass counter-terms by restoring the validity of chiral Ward identities of the massless theory. The left panel is taken from ref.~\cite{deDivitiis:2013xla} where simulations have been performed by using (Twisted Mass) Wilson fermions (different colors correspond to different lattice spacings) while the right panel is taken from ref.~\cite{Blum:2010ym} where simulations have been performed with Domain Wall fermions. The plots show that the parameters $\Delta m_f^{cr}$ can be obtained with high numerical precision. Details on the exact definitions of the chiral Ward identities used in the two cases can be found in the cited papers.
}  
\end{center}
\end{figure}
In lattice theories with broken chirality, the calculation of $\left[M_{K^+}-M_{K^0}\right]^\lQED$ can be performed provided that the \emph{linear divergent} counter-terms $\Delta m_f^{cr}$ have been accurately tuned. This can be done as in the case of the isosymmetric critical masses by restoring the validity of chiral Ward identities of the massless theory, see Figure~\ref{fig:mcritical}.

\begin{figure}[!t]
\begin{center}
\includegraphics[width=0.28\textwidth]{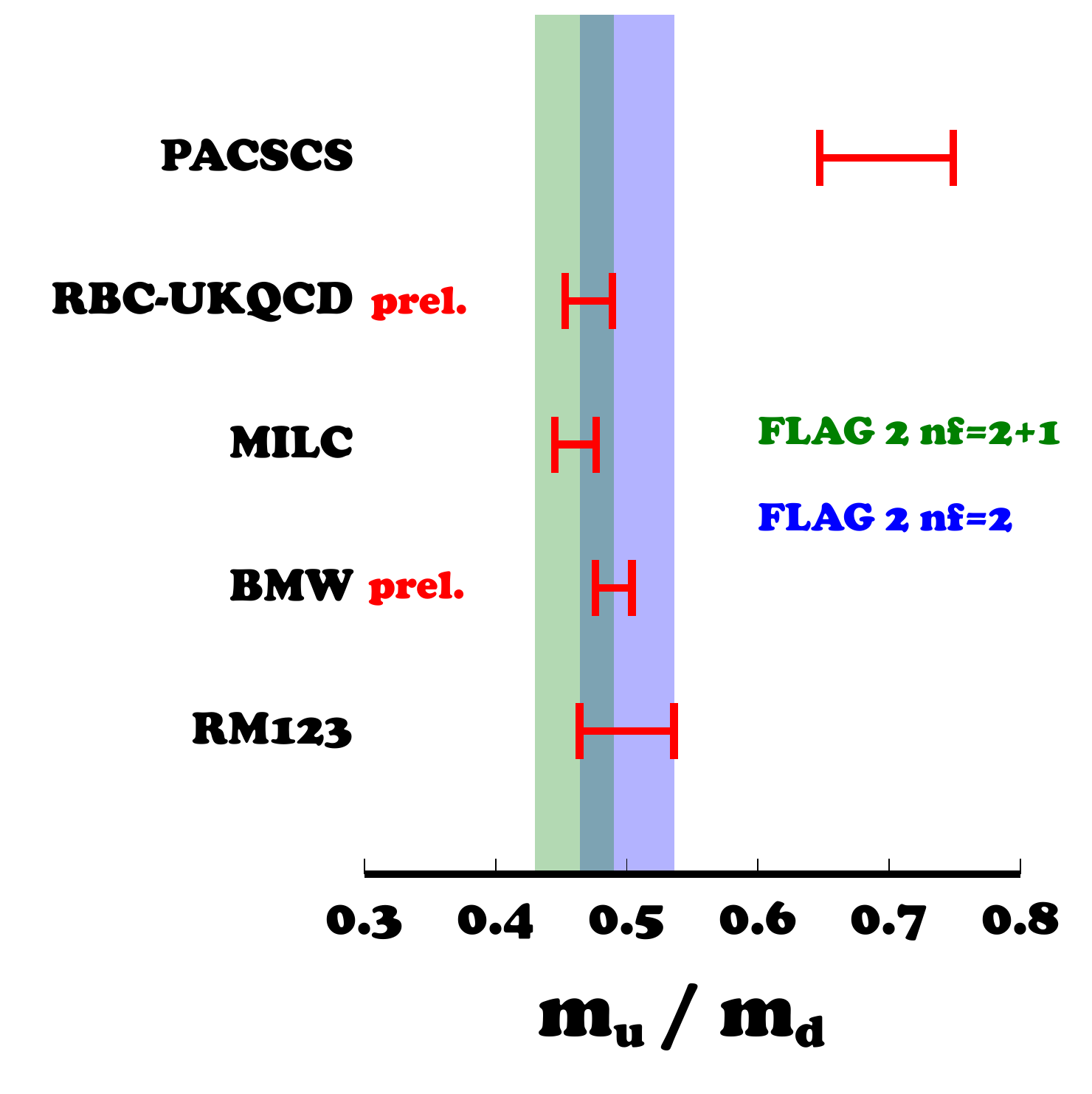} 
\qquad\qquad\qquad
\includegraphics[width=0.28\textwidth]{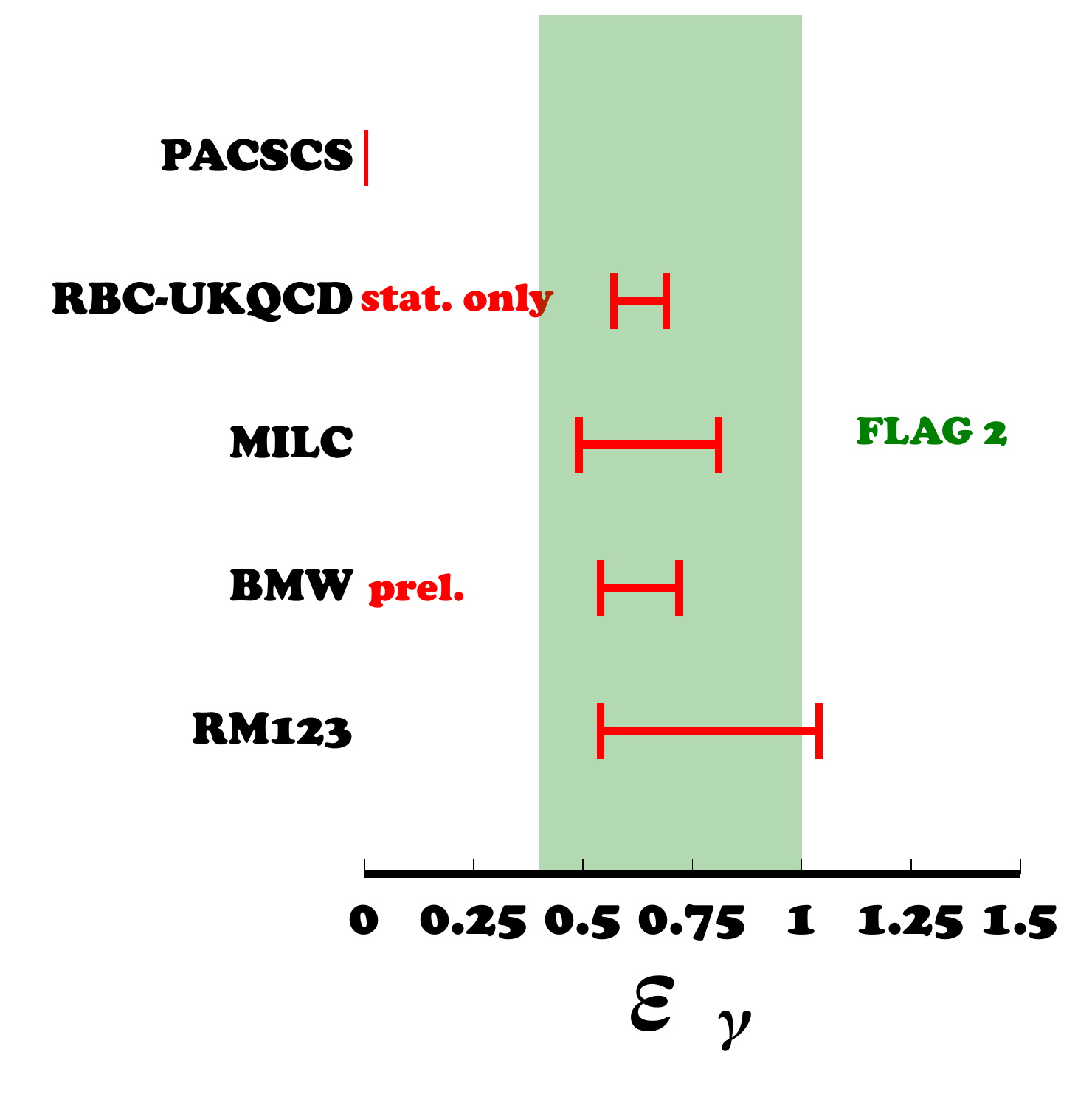}
\caption{\label{fig:epsilongamma} \footnotesize
Comparison plots of recent lattice determinations of $\hat m_u /\hat m_d$ (left) and $\varepsilon_\gamma$ (right). The green ($n_f=2+1$) and blue ($n_f=2$) bands represent the FLAG2 averages~\cite{Aoki:2013ldr} for these quantities.}  
\end{center}
\end{figure}
The results for $\left[M_{K^+}-M_{K^0}\right]^\lQED$ are usually expressed in terms of the Dashen's theorem breaking parameter $\varepsilon_\gamma$ (see ref.~\cite{Aoki:2013ldr} for the definition of other commonly used breaking parameters). The theorem follows from the observation that the electric charge operator is diagonal in flavour space: from the flavour vector symmetries of the full theory it follows $M_{K^+}=M_{\pi^+}+O(\hat m_s)$; from the flavour axial symmetries of the massless theory it follows that $M_{K^0}=O(\hat m_s)$ and $M_{\pi^0}=O(\hat m_{ud})$. The breaking parameter $\varepsilon_\gamma$ is  is a measure of the $O(\hat m_s)$ deviation from the chiral relation
\begin{eqnarray}
\hat m_s=\hat m_d=\hat m_u=0 \quad \mapsto \quad
\left[M_{K^+}-M_{K^0}\right]^\lQED=\left[M_{\pi^+}-M_{\pi^0}\right]^\lQED=\left[M_{\pi^+}-M_{\pi^0}\right]^{phys} 
\end{eqnarray}
and is defined as
\begin{eqnarray}
  \varepsilon_\gamma = 
  \frac{\left[M_{K^+}^2-M_{K^0}^2\right]^{QED}-
  \left[M_{\pi^+}^2-M_{\pi^0}^2\right]^{QED}}
  {M_{\pi^+}^2-M_{\pi^0}^2}
  =
  \frac{\left[M_{K^+}^2-M_{K^0}^2\right]^{QED}}
  {\left[M_{\pi^+}^2-M_{\pi^0}^2\right]^{QED}}
  -1  \ + O\left[ \Delta \hat m_{ud}\ \hat\alpha_{em} \right] \; .
\end{eqnarray}
Figure~\ref{fig:epsilongamma} shows the results obtained by the different collaborations for $\hat m_u/\hat m_d$ and $\varepsilon_\gamma$.
Note that in QCD+QED the ratio of quark masses is scale and scheme dependent and the results are given in the $\overline{MS}$ scheme at $\mu=\mu^\star=2$~GeV. Also the results for $\varepsilon_\gamma$ depend (mildly) on the renormalization prescriptions.
The RM123 results~\cite{deDivitiis:2013xla} have been obtained by the matching procedure discussed in this talk. The preliminary results~\cite{portellilattice} of the BMW collaboration have been obtained by using a matching procedure briefly discussed in ref.~\cite{Borsanyi:2013lga} (see also ref.~\cite{Portelli:2013jla}). The preliminary results~\cite{drurylattice} of the RBC-UKQCD collaboration (update of ref.~\cite{Blum:2010ym}) and of the MILC collaboration~\cite{duglattice} (update of ref.~\cite{Basak:2013iw}) have been obtained by using a renormalization prescription to separate QCD from QED IBE based on chiral perturbation theory fits of lattice data. The result of the PACS-CS collaboration has been obtained in ref.~\cite{Aoki:2012st}.

\section{Finite volume effects}
\label{sec:fve}
\begin{figure}[!t]
\begin{center}
\includegraphics[width=0.32\textwidth]{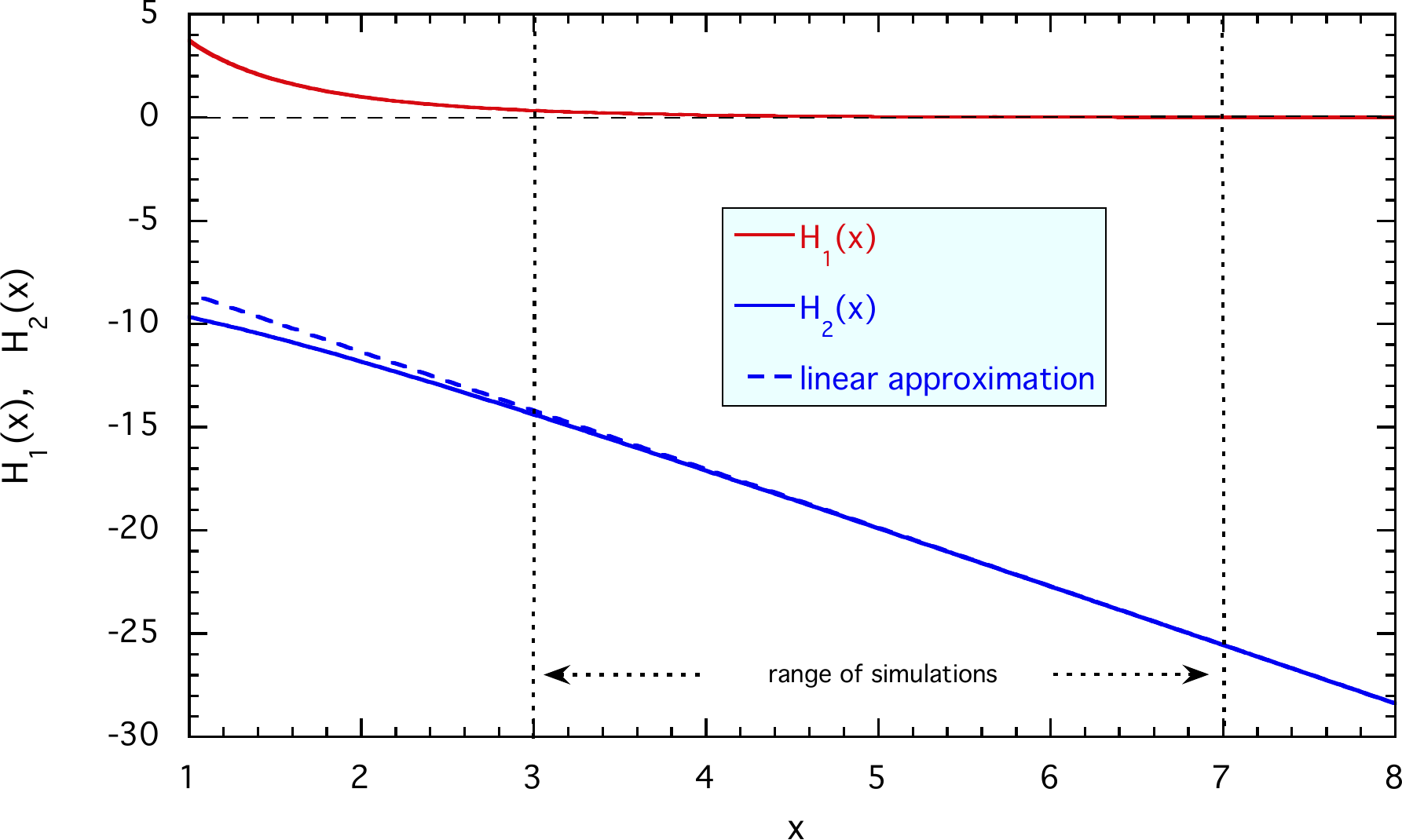} \hfill
\includegraphics[width=0.32\textwidth]{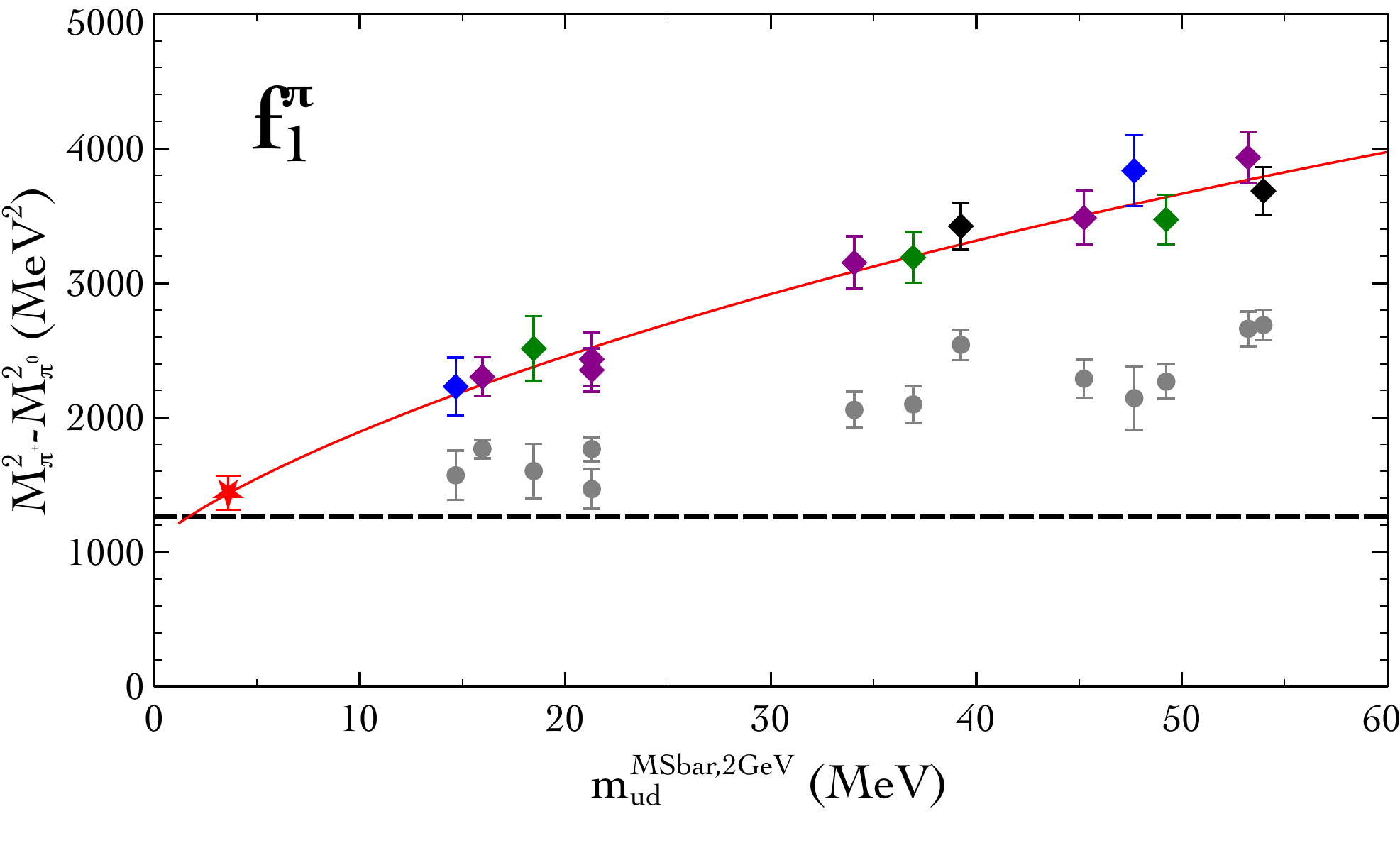} \hfill
\includegraphics[width=0.32\textwidth]{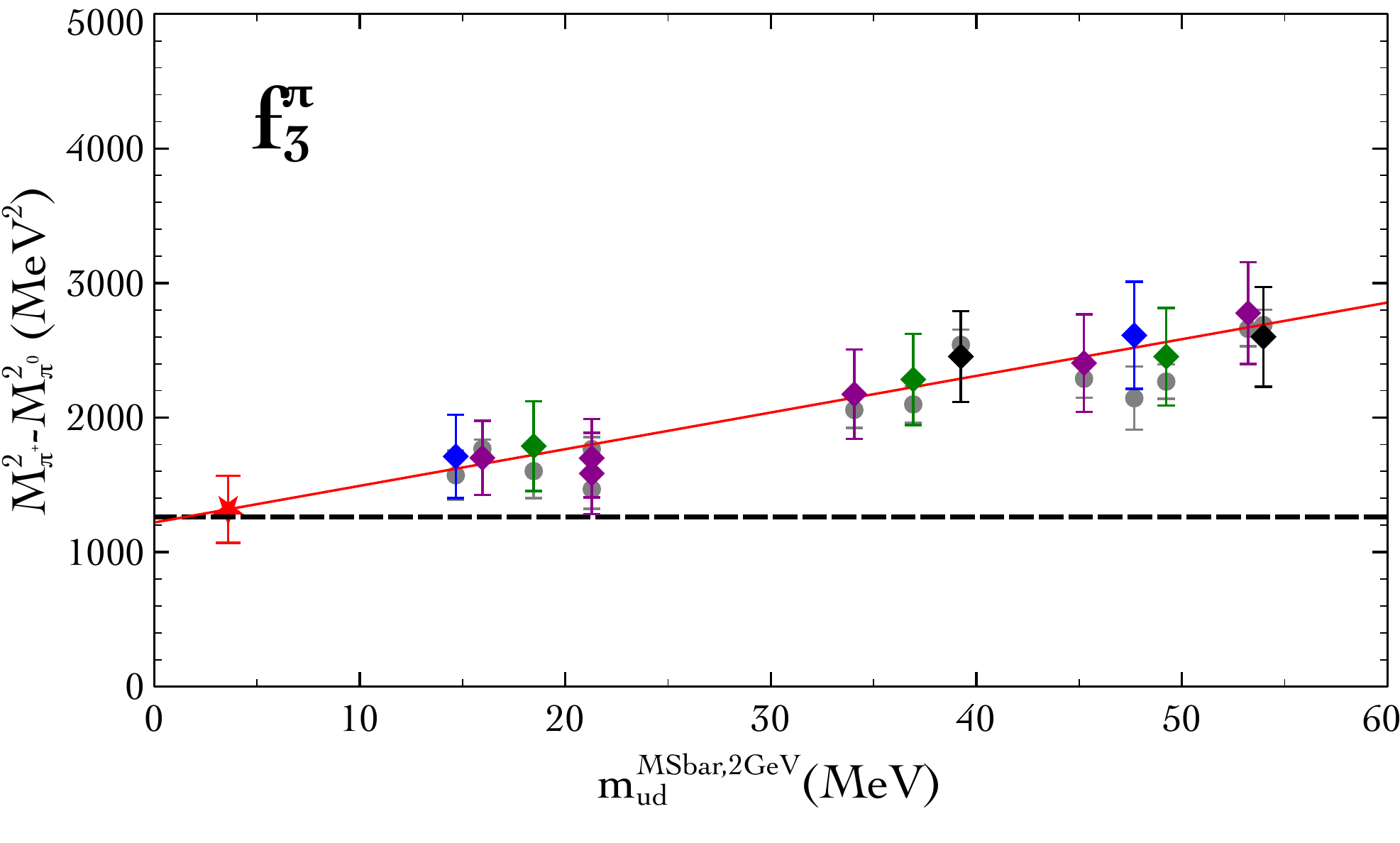}
\caption{\label{fig:pionfve} \footnotesize
Left panel: functions $H_{1,2}(x=M_\pi L)$ plotted together with their asymptotic expansions derived in ref.~\cite{deDivitiis:2013xla}. Center and right panels: combined chiral, continuum and infinite volume extrapolation of the pion mass splitting results of.~\cite{deDivitiis:2013xla}. In the center panel the chiral and infinite volume extrapolations are performed by using the chiral formulae of ref.~\cite{Hayakawa:2008an}. In the right panel the chiral and infinite volume extrapolations are performed by using a fitting function that depends linearly w.r.t. $\hat m_{ud}$ and $1/L^2$.
}  
\end{center}
\end{figure}
By putting photons in a box it is reasonable to expect large finite volume effects (FVE). This is presumably the main issue associated with lattice simulations of QCD+QED. In the case of light pseudoscalar meson masses, FVE have been calculated in chiral perturbation theory coupled to electromagnetism in ref.~\cite{Hayakawa:2008an}. For the pion mass splitting one gets
\begin{eqnarray}
\left[M_{\pi^+}^2-M_{\pi^0}^2\right](L)-\left[M_{\pi^+}^2-M_{\pi^0}^2\right](\infty)
&=&
\frac{ \hat e^2}{4\pi L^2} \left[
H_2(M_\pi L)-4C H_1(M_\pi L)
\right]
\nonumber \\
&\sim& 
-\frac{ \hat e^2 2.8373\dots}{4\pi}\left(\frac{M_\pi}{L}+\frac{2}{L^2}\right) \; ,
\end{eqnarray}
where the functions $H_{1,2}(x)$ are plotted in the left panel of Figure~\ref{fig:pionfve}. Similar results have been obtained for the kaon mass splitting. According to the previous expression, leading FVE go as $M_{\pi}/L$ and/or as $1/L^2$ and may be a as large as $30$\%. In ref.~\cite{deDivitiis:2013xla} these formulae have been used to fit the lattice data for the pion mass splitting previously shown in Figure~\ref{fig:ourslopes}. The fit is shown in the center panel of Figure~\ref{fig:pionfve}: the effect of the finite volume correction (difference between grey and coloured points) is somehow balanced by the chiral-log curvature and, within the errors, the final result is compatible with the experimental value of $M_{\pi^+}^2-M_{\pi^0}^2$ (black dashed line). In the right panel of Figure~\ref{fig:pionfve} the same lattice data are extrapolated by using a phenomenological fitting function, linear in $\hat m_{ud}$ and $1/L^2$: in this case the fitted FVE are much smaller than the chiral perturbation theory prediction and the final result is again compatible with the experimental determination. Both the fits of Figure~\ref{fig:pionfve} come with $\chi^2/dof \sim 1$.

\begin{figure}[!t]
\begin{center}
\includegraphics[width=0.3\textwidth]{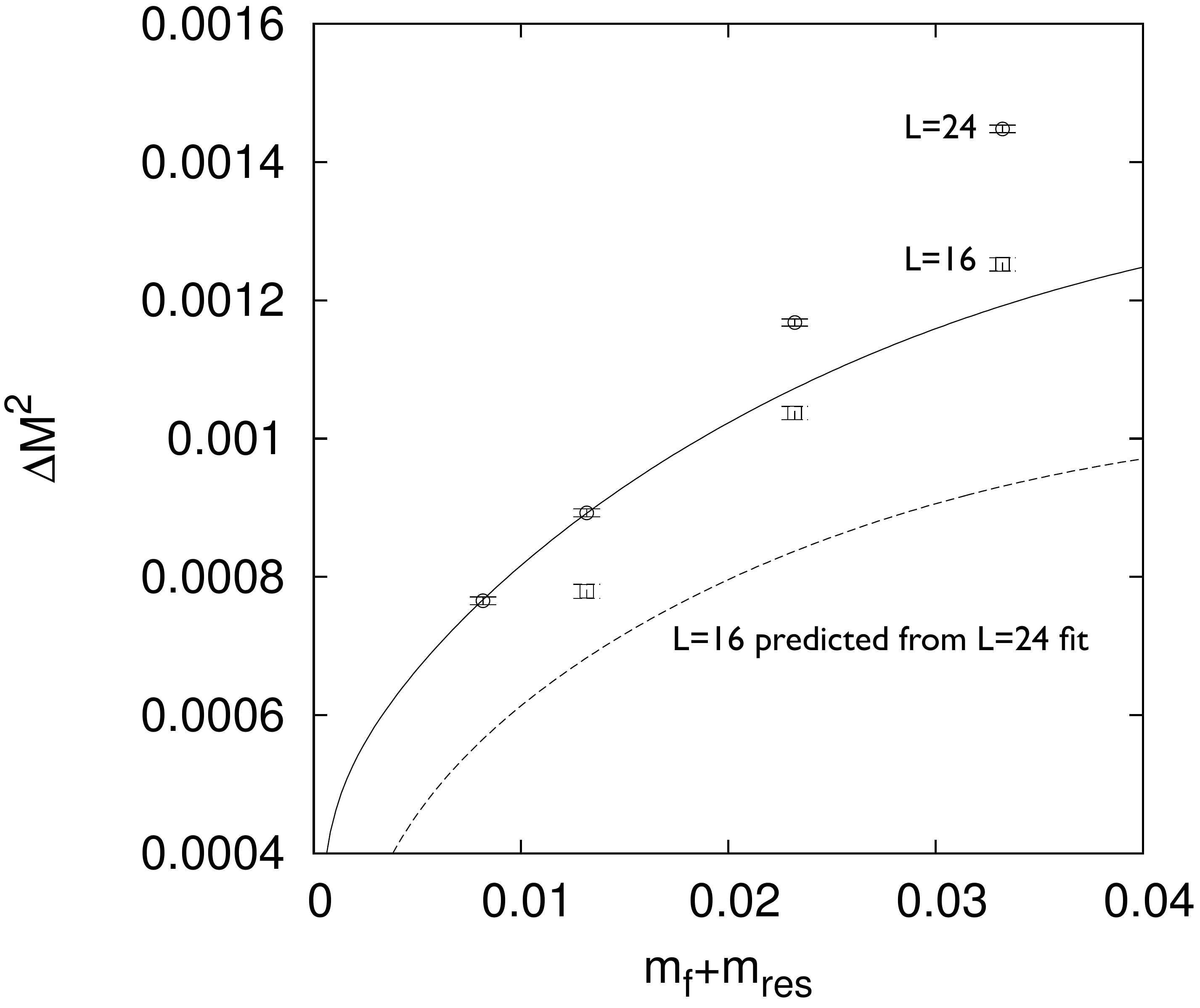} \hfill
\includegraphics[width=0.26\textwidth]{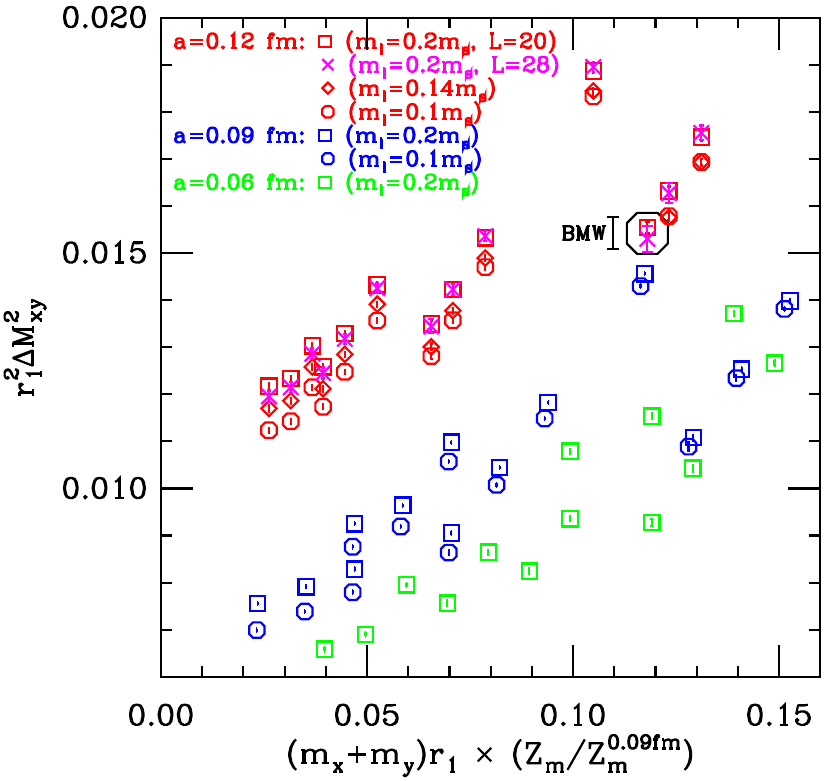} \hfill
\includegraphics[width=0.36\textwidth]{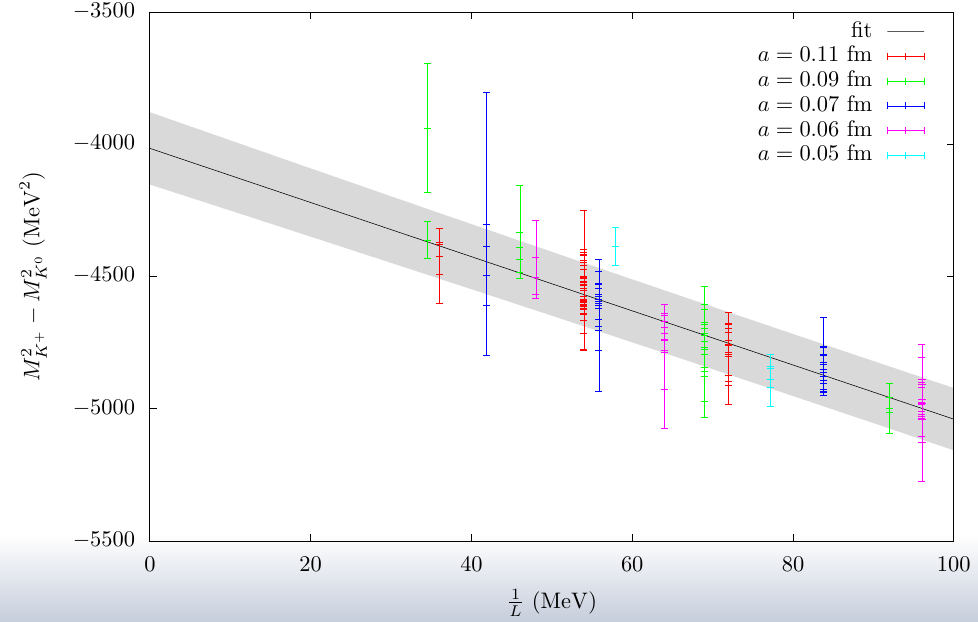}
\caption{\label{fig:fve} \footnotesize
Finite volume effects on the light pseudoscalar meson masses obtained by the UKQCD-RBC collaboration~\cite{Blum:2010ym} (left panel), the MILC collaboration~\cite{Basak:2013iw} (center panel) and by the BMW collaboration~\cite{Portelli:2013jla,portellilattice} (right panel).
}  
\end{center}
\end{figure}
Similar results have been found by other groups. Figure~\ref{fig:fve} shows the results of the RBC-UKQCD collaboration~\cite{Blum:2010ym} (left panel), of the MILC collaboration~\cite{Basak:2013iw} (center panel) and of the BMW collaboration~\cite{Portelli:2013jla,portellilattice} (right panel). The RBC-UKQCD collaboration used the FVE chiral formulae of ref.~\cite{Hayakawa:2008an} to fit the data obtained on a volume with $aL=24$ ($L\sim 3$~fm). The results of this fit have then been used to ``predict'' the data obtained on a smaller physical volume ($aL=16$) and a sizeable discrepancy has been observed. The MILC collaboration results also suggest that measured FVE may be much smaller than the ones predicted in chiral perturbation theory.
The BMW collaboration has obtained results on several gauge ensembles, including simulations at the physical pion mass and on volumes as large as $L=6$~fm. The right panel of Figure~\ref{fig:fve} shows the infinite volume extrapolation of the BMW (preliminary) results performed by parametrizing FVE with a term proportional to $1/L$. The resulting FVE are of the same order of magnitude of the chiral perturbation theory results. In summary, given the size of the statistical and other systematic errors on the lattice results for pseudoscalar meson masses, it is not possible to establish at present if the measured finite volume effects confirm the chiral perturbation theory predictions. 

\begin{figure}[!t]
\begin{center}
\includegraphics[width=0.36\textwidth]{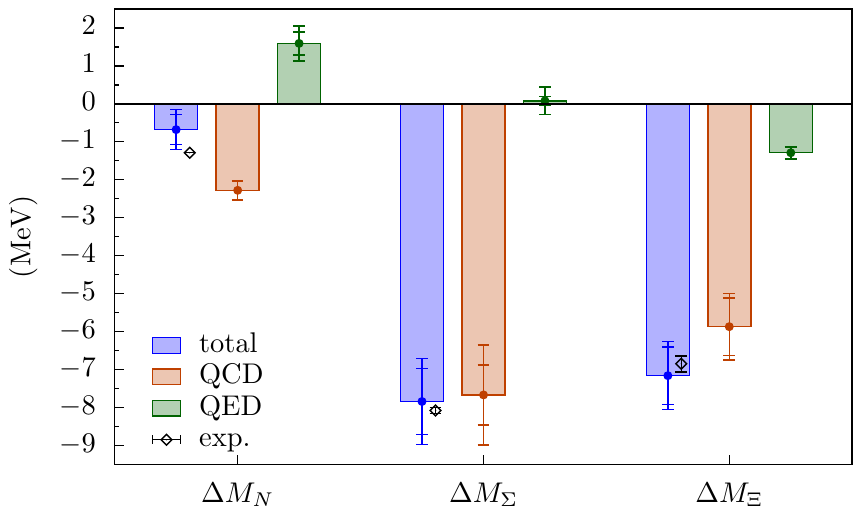} \hfill
\includegraphics[width=0.36\textwidth]{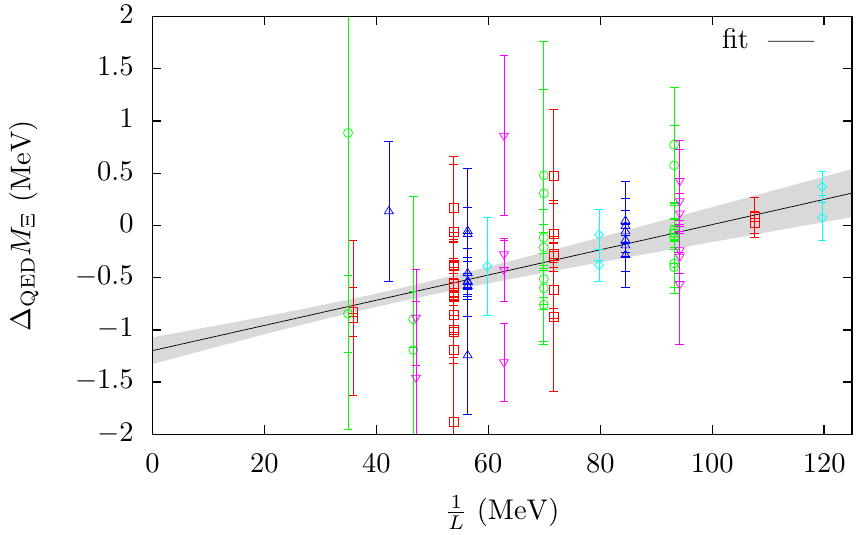} \hfill
\includegraphics[width=0.25\textwidth]{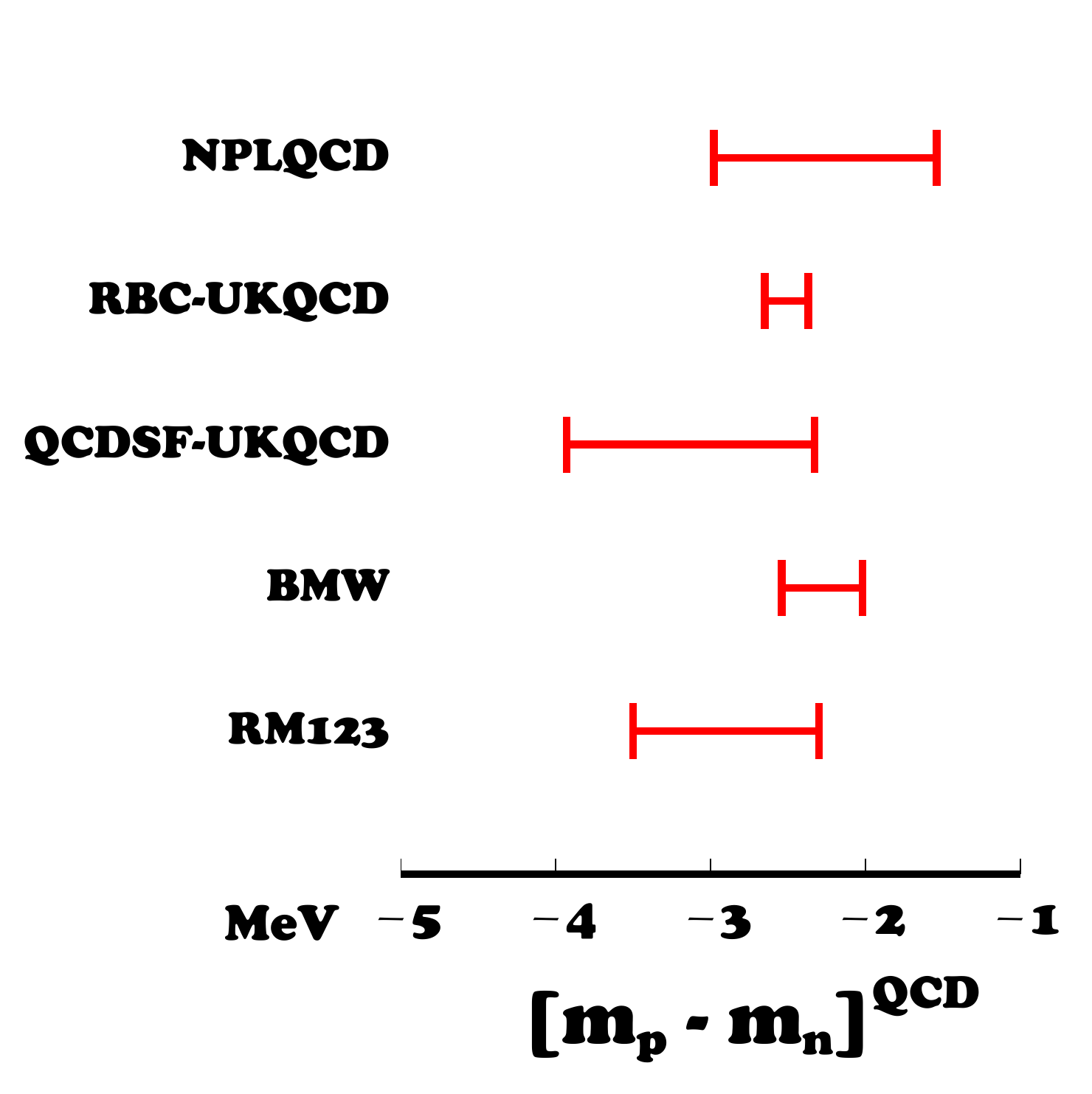}
\caption{\label{fig:baryons} \footnotesize
IBE effects on the octet baryon masses obtained by the BMW collaboration (left and center panel). In the right panel is shown a comparison plot of the results obtained by the different collaborations for the QCD contribution to the proton-neutron mass splitting.
}  
\end{center}
\end{figure}
The BMW collaboration has recently completed~\cite{Borsanyi:2013lga} a systematic investigation of IBE on the octet baryon masses. The results for the QED, QCD and total contributions to the mass splittings are shown in the left panel of Figure~\ref{fig:baryons}. In the center panel of the Figure the BMW results are fitted linearly in $1/L$. The statistical errors are still very large but the fit shows that FVE on baryon masses can be as large as $80$\%! The right panel of the Figure shows a comparison plot of the results obtained by the different collaborations for the QCD contribution to the proton-neutron mass splitting. The NPLQCD result has been obtained in ref.~\cite{Beane:2006fk}, the RBC-UKQCD result in ref.~\cite{Blum:2010ym}, the QCDSF-UKQCD result in ref.~\cite{Horsley:2012fw} and the RM123 result in ref.~\cite{deDivitiis:2011eh}. There is a substantial agreement between the different determinations and, by relying in particular on the BMW result, this is a first confirmation that the proton cannot decay weakly.  

\section{IBE on hadronic matrix elements}
In this last section I want to briefly discuss the problem of the calculation of LIBE in hadronic processes, for example in the $K_{\ell2}$ decay rate. The physical observable in this case is $\Gamma[K^+\mapsto \ell^+ \nu (\gamma)]$, including soft photons. This is ultraviolet and infrared finite, gauge invariant, unambiguous. Because of the presence of contributions as the one shown in Figure~\ref{fig:exchange} the decay rate cannot be factored into an hadronic and a leptonic part and it can be misleading to talk about $F_K$ without specifying further details (see ref.~\cite{Gasser:2010wz} for a discussion of this point in the framework of chiral perturbation theory). 

\begin{figure}[!t]
\begin{center}
\includegraphics[width=0.25\textwidth]{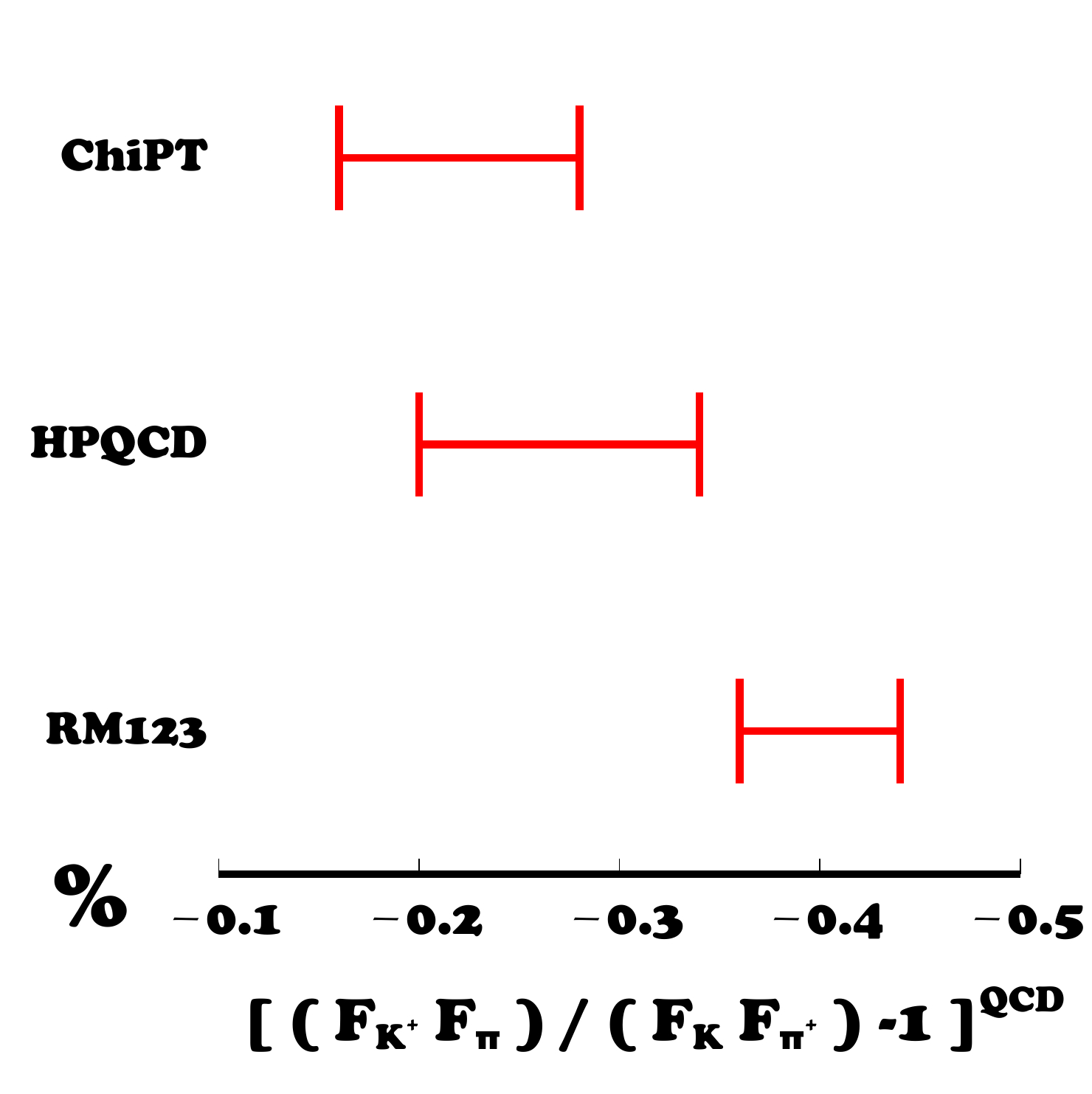}
\caption{\label{fig:fkfpi} \footnotesize
Comparison of the RM123 and HPQCD lattice results for the QCD IBE on the ratio $F_K/F_\pi$ with the chiral perturbation theory result of ref.~\cite{Cirigliano:2011tm}.
}  
\end{center}
\end{figure}
On the other hand, by specifying a prescription to separate QED from QCD IBE effects, the QCD corrections can be properly defined and accurately calculated on the lattice. This is the approach followed in ref.~\cite{deDivitiis:2011eh} where QCD IBE corrections to the ratio $F_K/F_\pi$ have been calculated by starting from eq.~(\ref{eq:QCDkaons}). Similar results have been obtained in ref.~\cite{Dowdall:2013rya} where leading QCD IBE on the kaon decay constant have been calculated by starting from correlators with $m_u\neq m_d$ and by relying on chiral perturbation theory. The two lattice results are compared with the chiral perturbation theory calculation of ref.~\cite{Cirigliano:2011tm} in Figure~\ref{fig:fkfpi}: lattice data confirm that QCD IBE on the $K_{\ell2}$ decay rate are of the order of a few permille, i.e. comparable with the overall uncertainty quoted on $F_K/F_\pi$ in ref.~\cite{Aoki:2013ldr}. A detailed discussion of the theoretical issues associated with a first principle calculation of the QCD+QED IBE corrections to the decay rate will be the subject of ref.~\cite{inprep}.

\section{Conclusions}
Isospin breaking effects can be calculated on the lattice from first principles, even including QED unquenching effects. QCD+QED observables can be evaluated by starting from isosymmetric QCD lattice simulations using reweighting techniques. On volumes $L\sim 3$~fm it has been demonstrated that the fluctuations of the reweighting factor can be kept under control. By simulating the full theory at the physical values of the parameters $\hat m_d-\hat m_u$ and $\hat \alpha_{em}$ it is difficult to extract IBE because, in general, these are smaller than the statistical errors.
Leading isospin breaking effects can also be obtained by expanding the relevant correlators with respect to the up-down mass difference and the electric charge. This approach allows to obtain large numerical signals but it may require the calculation of several correlators. 

Finite volume effects are the main issue. This is not surprising, lattice simulations have to be performed on a finite volume and QED is a long-range unconfined interaction. On pseudoscalar meson masses FVE can be as large as $30$\% and even larger on baryon masses. Although this is a potentially very large systematic error, we are nowadays \emph{calculating}, not just \emph{guessing}, isospin breaking effects. Even a large uncertainty on isospin breaking effects is a small and reliable uncertainty on the given observable: $1\% \times 30\% = 0.3\%$! 

\section*{Acknowledgements}
I warmly thank my colleagues of the RM123 collaboration for the enjoyable and fruitful work on the subjects covered in this talk. In particular I thank V.~Lubicz for his comments on this manuscript.

\end{document}

%% file: newcommand.tex
\newcommand{\sla}[1]%
        {\kern .25em\raise.18ex\hbox{$/$}\kern-.6em #1}

\newcommand{\goi}{\raisebox{-0.3\totalheight}{\includegraphics[scale=.3]{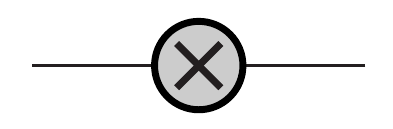}}}

\newcommand{\golself}{\raisebox{-0.15\totalheight}{\includegraphics[scale=.3]{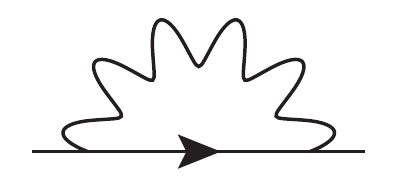}}}

\newcommand{\gdll}{\raisebox{-0.4\totalheight}{\includegraphics[scale=.18]{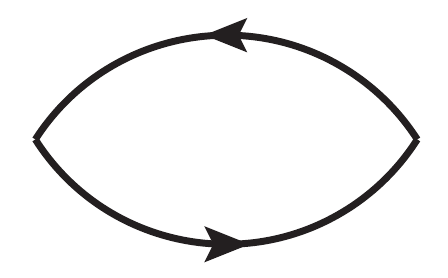}}}
\newcommand{\gdsi}{\raisebox{-0.4\totalheight}{\includegraphics[scale=.18]{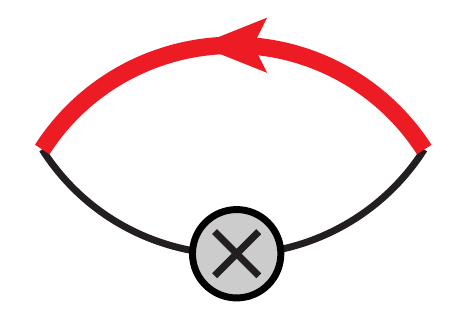}}}
\newcommand{\gdsip}{\raisebox{-0.4\totalheight}{\includegraphics[scale=.18]{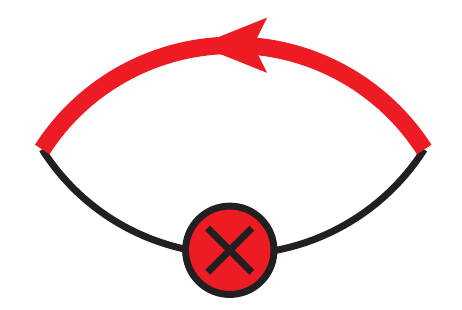}}}

\newcommand{\gdsl}{\raisebox{-0.4\totalheight}{\includegraphics[scale=.18]{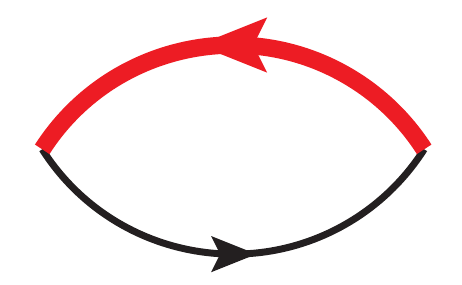}}}

\newcommand{\gdslselfl}{\raisebox{-0.6\totalheight}{\includegraphics[scale=.18]{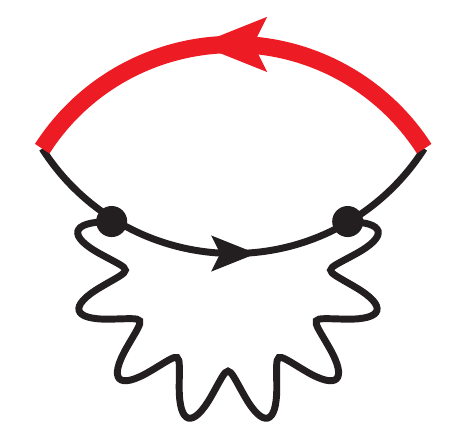}}}
\newcommand{\gdslexch}{\raisebox{-0.4\totalheight}{\includegraphics[scale=.18]{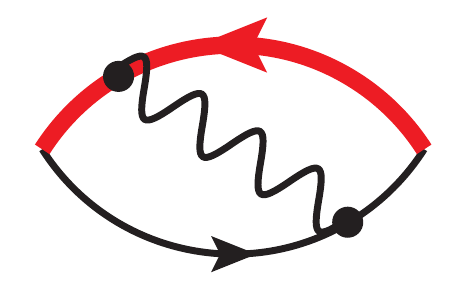}}}

\newcommand{\gdllexch}{\raisebox{-0.4\totalheight}{\includegraphics[scale=.18]{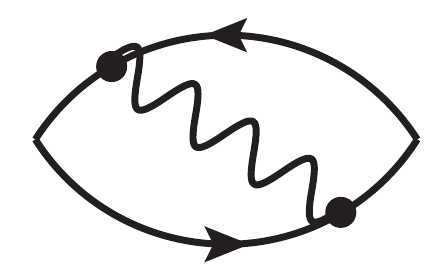}}}
\newcommand{\discgdllexch}{\raisebox{-0.2\totalheight}{\includegraphics[scale=.3]{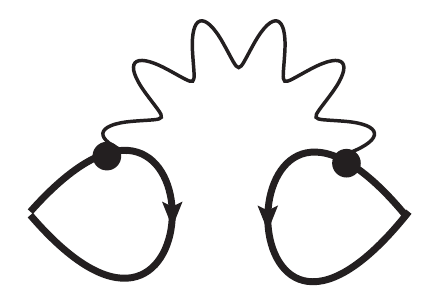}}}

\newcommand{\gdslltadf}{\raisebox{-0.4\totalheight}{\includegraphics[scale=.18]{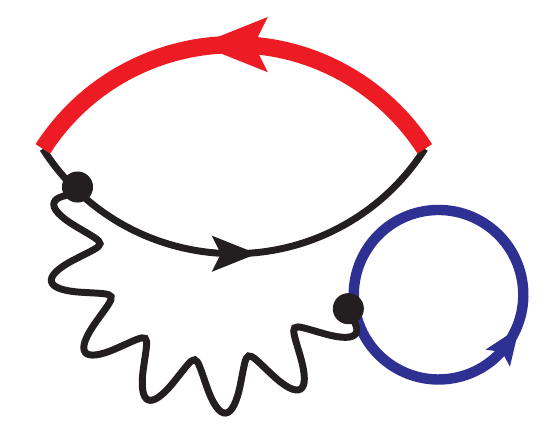}}}

\newcommand{\gdslphtadl}{\raisebox{-0.6\totalheight}{\includegraphics[scale=.18]{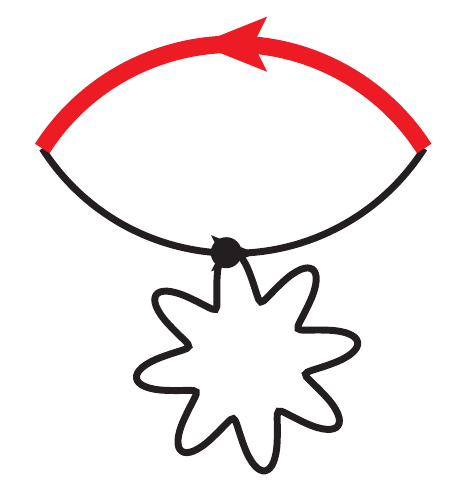}}}

\newcommand{\bear}[1]{\begin{equation}\begin{array}{#1}}
\newcommand{\eear}{ \end{array}\end{equation}}
\newcommand{\barr}[1]{\begin{array}{#1}}
\newcommand{\earr}{\end{array}}
